\documentclass[apj]{emulateapj}
\slugcomment{}

\shorttitle{Rates and Properties of Type Ia Supernovae}
\shortauthors{Sullivan et al.}

\newcommand\omatter{\ifmmode \Omega_{M}\else $\Omega_{M}$\fi}
\newcommand\ok{\ifmmode \Omega_{\mathrm{k}}\else $\Omega_{\mathrm{k}}$\fi}
\newcommand\olambda{\ifmmode \Omega_{\Lambda}\else $\Omega_{\Lambda}$\fi}
\newcommand\dmB{\ifmmode \Delta m_{15}(B) \else $\Delta m_{15}(B)$\fi}
\newcommand\zspec{\ifmmode z_{\mathrm{spec}}\else $z_{\mathrm{spec}}$\fi}
\newcommand\zphot{\ifmmode z_{\mathrm{phot}}\else $z_{\mathrm{phot}}$\fi}
\newcommand\deltaz{\ifmmode \Delta z \else $\Delta z$\fi}
\newcommand\chidof{\ifmmode \chi^2/\mathrm{DOF}\else $\chi^2/\mathrm{DOF}$\fi}
\newcommand\ebmvmw{\ifmmode E_{B-V}^{\mathrm{mw}}\else $E_{B-V}^{\mathrm{mw}}$\fi}
\newcommand\ebmvhost{\ifmmode E_{B-V}^{\mathrm{host}}\else $E_{B-V}^{\mathrm{host}}$\fi}
\newcommand\mperyr{\ifmmode \mathrm{M}_{\odot}\,\mathrm{yr}^{-1} \else $\mathrm{M}_{\odot}\,\mathrm{yr}^{-1}$ \fi}
\newcommand\vmax{\ifmmode V_{\mathrm{max}}\else $V_{\mathrm{max}}$\fi}
\newcommand\ergshz{\ifmmode \mathrm{erg}\,\mathrm{s}^{-1}\,\mathrm{Hz}^{-1}\else $\mathrm{erg}\,\mathrm{s}^{-1}\,\mathrm{Hz}^{-1}$ \fi}
\newcommand\msun{\ifmmode \mathrm{M}_{\odot} \else $\mathrm{M}_{\odot}$\fi}
\newcommand\snrIa{\ifmmode \mathrm{SNR}_{\mathrm{Ia}} \else $\mathrm{SNR}_{\mathrm{Ia}}$\fi}
\newcommand\Aunits{\ifmmode \mathrm{SNe}\,\,\mathrm{yr}^{-1}\,\msun^{-1} \else $\mathrm{SNe}\,\,\mathrm{yr}^{-1}\,\msun^{-1}$ \fi}
\newcommand\Bunits{\ifmmode \mathrm{SNe}\,\,\mathrm{yr}^{-1}\,(\mperyr)^{-1} \else $\mathrm{SNe}\,\,\mathrm{yr}^{-1}\,(\mperyr)^{-1}$ \fi}

\begin{document}


\title{Rates and properties of type Ia supernovae as a function of
  mass and star-formation in their host galaxies}

\author{M.~Sullivan\altaffilmark{1}, D.~Le~Borgne\altaffilmark{2}, C.~J.~Pritchet\altaffilmark{3}, A.~Hodsman\altaffilmark{1}, J.~D.~Neill\altaffilmark{3}, D.~A.~Howell\altaffilmark{1}, R.~G.~Carlberg\altaffilmark{1}, P.~Astier\altaffilmark{4}, E.~Aubourg\altaffilmark{5,2}, D.~Balam\altaffilmark{3}, S.~Basa\altaffilmark{6}, A.~Conley\altaffilmark{1}, S.~Fabbro\altaffilmark{7}, D.~Fouchez\altaffilmark{8}, J.~Guy\altaffilmark{4}, I.~Hook\altaffilmark{9}, R.~Pain\altaffilmark{4}, N.~Palanque-Delabrouille\altaffilmark{2}, K.~Perrett\altaffilmark{1}, N.~Regnault\altaffilmark{4}, J.~Rich\altaffilmark{2}, R.~Taillet\altaffilmark{10,4}, S.~Baumont\altaffilmark{4}, J.~Bronder\altaffilmark{9}, R.~S.~Ellis\altaffilmark{11}, M.~Filiol\altaffilmark{6}, V.~Lusset\altaffilmark{2}, S.~Perlmutter\altaffilmark{12}, P.~Ripoche\altaffilmark{8}, C.~Tao\altaffilmark{8}}

\altaffiltext{1}{Department of Astronomy and Astrophysics, University
  of Toronto, 60 St. George Street, Toronto, ON M5S 3H8, Canada}
\altaffiltext{2}{DAPNIA/Service d'Astrophysique, CEA-Saclay, 91191 Gif-sur-Yvette Cedex, France} 
\altaffiltext{3}{Department of Physics and Astronomy, University of
  Victoria, PO Box 3055, Victoria, BC V8W 3P6, Canada}
\altaffiltext{4}{LPNHE, CNRS-IN2P3 and University of Paris
  VI \& VII, 75005 Paris, France} 
\altaffiltext{5}{APC, 11 Pl. M.
  Berthelot, 75231 Paris Cedex 5, France}
\altaffiltext{6}{LAM CNRS,
  BP8, Traverse du Siphon, 13376 Marseille Cedex 12, France}
\altaffiltext{7}{CENTRA - Centro Multidisciplinar de Astrof\'{\i}sica,
  IST, Avenida Rovisco Pais, 1049 Lisbon, Portugal}
\altaffiltext{8}{CPPM, CNRS-IN2P3 and University Aix Marseille II,
  Case 907, 13288 Marseille Cedex 9, France}
\altaffiltext{9}{University of Oxford Astrophysics, Denys Wilkinson
  Building, Keble Road, Oxford OX1 3RH, UK}
\altaffiltext{10}{Université de Savoie, 73000 Chamb\'ery, France}
\altaffiltext{11}{California Institute of Technology, E. California
  Blvd., Pasadena, CA 91125, USA}
\altaffiltext{12}{Lawrence Berkeley National Laboratory, 1 Cyclotron
  Rd., Berkeley, CA 94720, USA}

\begin{abstract}
  
  We show that Type Ia supernovae (SNe~Ia) are formed within both very
  young and old stellar populations, with observed rates that depend
  on the stellar mass and mean star-formation rates (SFRs) of their
  host galaxies.  Models where the SN~Ia rate depends solely on host
  galaxy stellar mass are ruled out with $>$99\% confidence.  Our
  analysis is based on 100 spectroscopically-confirmed SNe~Ia, plus 24
  photometrically-classified events, all from the Supernova Legacy
  Survey (SNLS) and distributed over 0.2$<$z$<$0.75.  Using multi-band
  photometry, we estimate stellar masses and SFRs for the SN~Ia host
  galaxies by fitting their broad-band spectral energy distributions
  with the galaxy spectral synthesis code, PEGASE.2.  We show that the
  SN~Ia rate per unit mass is proportional to the specific SFR of the
  parent galaxies -- more vigorously star-forming galaxies host more
  SNe~Ia per unit stellar mass, broadly equivalent to the trend of
  increasing SN~Ia rate in later-type galaxies seen in the local
  universe.  Following earlier suggestions for a simple
  ``two-component'' model approximating the SN~Ia rate, we find
  bivariate linear dependencies of the SN~Ia rate on both the stellar
  masses and the mean SFRs of the host systems. We find that the SN~Ia
  rate can be well represented as the sum of $5.3\pm1.1\times10^{-14}$
  SNe per year per unit stellar mass, and $3.9\pm0.7\times10^{-4}$ SNe
  per year per $\msun\,\mathrm{yr}^{-1}$ of star formation.
  
  We also demonstrate a dependence of distant SN~Ia light-curve shapes
  on star-formation in the host galaxy, similar to trends observed
  locally. Passive galaxies, with no star-formation, preferentially
  host faster-declining/dimmer SNe~Ia, while slower-declining/brighter
  events are only found in systems with ongoing star-formation.  We
  model the light-curve width distribution in star-forming galaxies as
  the sum of a young component, and an old component taken from the
  distribution in non-star-forming galaxies.  Empirically
  understanding these relationships between SNe~Ia and their
  environments will lead to future improvements in their use as
  cosmological candles.

\end{abstract}



\section{Introduction}
\label{sec:introduction}

Type Ia supernovae (SNe~Ia) represent cosmologists' most direct probe
of the cosmic expansion history, yet an understanding of the
composition of their progenitor systems has not yet been achieved
\citep[e.g.][]{2000ARA&A..38..191H}. In principle, this uncertainty
can be reduced and constraints placed on the nature of the progenitor
if the typical explosion time-scale of SNe~Ia can be determined. This
``delay-time'' parameterizes the distribution of times between the
binary system formation and subsequent SN explosion following
accretion of material from a secondary companion \citep*[e.g.][see
\citealt{2005A&A...441.1055G} for a review of current rate model
formalizations]{1998MNRAS.297L..17M}.

Various pieces of observational evidence have been used to place
different constraints on the value of this delay time. The principal
approach compares observed SN~Ia rates with a predicted rate generated
by convolving a delay function with an assumed cosmic star-formation
history \citep[e.g.][]{1998MNRAS.297L..17M,2004MNRAS.347..942G}.
Studies using this approach have determined a wide range of
delay-times: $\simeq$2-4\,Gyr
\citep{2004ApJ...613..200S,2005ApJ...635.1370S}, $\ge2$\,Gyr
\citep{2004MNRAS.347..942G}, and $\le1$\,Gyr
\citep{2006ApJ...637..427B}.

A second technique uses comparisons of the SN~Ia rate in $z<1$ galaxy
clusters with the observed cluster iron content. If SNe~Ia are assumed
to represent the dominant source of iron in clusters, the low cluster
SN~Ia rate at low redshift implies much of the iron must have been
produced from events at higher redshift -- and hence suggests delay
times of $<$2\,Gyr \citep{2004MNRAS.347..951M}.  The third piece of
evidence follows from the observation that SNe~Ia are substantially
more common in star-forming later-type galaxies than in early-type
systems \citep[e.g.][]{1979AJ.....84..985O}.  The rate per unit mass
is significantly higher both in later-type galaxies than in E/S0
systems
\citep{1990PASP..102.1318V,1994ApJ...423L..31D,2005A&A...433..807M},
and in bluer galaxies than in red galaxies
\citep{2005A&A...433..807M}, with an enhancement of SNe~Ia in
early-type galaxies that are radio-loud \citep{2005ApJ...629..750D}.
Furthermore, within star-forming galaxies, SNe~Ia are rarer in galaxy
bulges than in their discs \citep*{1997ApJ...483L..29W}. This suggests
some dependence of the SN~Ia rate on recent star-formation (and hence
very short delay-times).  Finally, there is the observation that
SNe~Ia are common in old evolved systems with little recent
star-formation activity
\citep{1999A&A...351..459C,2005A&A...433..807M}, suggesting some
progenitors have delay-times of at least several Gyr; other studies of
the star-formation histories of local SN~Ia host galaxies claim a
delay-time lower limit of $\sim$2\,Gyr \citep{2005ApJ...634..210G}.

These contradictions can be resolved by removing the constraint of a
single delay-time parameterizing all SN~Ia explosions, and instead
using a ``two component'' distribution (or even a more general
function), similar to the models proposed by
\citet{2005A&A...433..807M,2005astro.ph.10315M} and \citet[hereafter
SB05]{2005ApJ...629L..85S}. Such models comprise a ``prompt'' (or
small delay-time) SN~Ia component, essentially dependent on recent
star-formation, and an ``old'' (larger delay-time) component,
dependent on the number of lower-mass stars. The total rate of SNe~Ia
is then a combination of these different functions. These scenarios
are able to resolve many of the observational contradictions described
above (see discussion in SB05).

These two-component models have implicit but important implications
for the progenitors of SNe~Ia which might impact their use as
calibrated standard candles to derive cosmological parameters
\citep{1998AJ....116.1009R,1999ApJ...517..565P,2006A&A...447...31A}.
The possibility of subtle differences between SNe~Ia from the two
components, or a change in the relative fraction of the two types with
redshift, are potentially concerning. As such it is vital to test and
parameterize the SN~Ia models in as many ways as possible. One way to
do this is to look at the rates and properties of SNe~Ia in relation
to the environments or galaxies in which they explode. Galaxies across
the Hubble sequence provide an ideal laboratory for studying SNe~Ia
due the range of potential progenitor stellar populations that can be
probed, from star-burst galaxies with dominant young populations of
stars, through normal galaxies such as the Milky Way with a
substantial fraction of evolved stellar mass, to the old, evolved
elliptical galaxies in massive galaxy clusters which are essentially
comprised of homogeneous old stellar populations.

In this paper, we examine the properties of high-redshift SN~Ia host
galaxies, and calculate the frequency of occurrence of SNe~Ia in
galaxies of different type, exploring the parameters governing the
SN~Ia rate. Our motivation is to measure the SN~Ia rate as a function
of the stellar mass and star-formation rates of the host galaxies in
order to test the various predictions of the models described above.
Performing these tests requires not only a large, homogeneous dataset
of SNe~Ia with well-understood detection efficiency characteristics,
but additionally a multi-wavelength dataset for their host galaxy
systems which can be used to constrain their stellar populations. The
intermediate-redshift Supernova Legacy Survey (SNLS) satisfies both
requirements.

The SNLS uses repeat $u^*g'r'i'z'$ imaging of four square-degree
fields to conduct a ``rolling'' high-redshift SN search.  The repeat
imaging allows not only the construction of high-quality multi-band
$g'r'i'z'$ SN light-curves \citep{2006A&A...447...31A}, but also
extremely deep images of the survey fields, and hence precise
information on the broad-band spectral energy distributions (SEDs) of
the SN host galaxies can be collected. Perhaps surprisingly, precision
work on SNe~Ia rates is most efficiently performed at these
intermediate redshifts ($z\sim0.5$); low-redshift SN~Ia rates are
still far less precise than those at higher redshift
\citep[e.g.][]{2006neill} due to the homogeneity of high-redshift
rolling SN searches.

A plan of the paper follows. $\S$~\ref{sec:snls-dataset},
$\S$~\ref{sec:galaxy-meas} and $\S$~\ref{sec:incompl-corr} detail the
framework upon which our subsequent analyses in
$\S$~\ref{sec:supernova-rate} and $\S$~\ref{sec:supern-prop-as} are
built. In $\S$~\ref{sec:snls-dataset} we introduce the SNLS and
discuss the construction of deep optical stacks of the survey field as
well as the host galaxy identification and flux measurement.
$\S$~\ref{sec:galaxy-meas} describes the galaxy SED fitting, and
$\S$~\ref{sec:incompl-corr} details the various incompleteness
corrections we apply to our galaxy and SN~Ia samples. Our analysis is
contained in $\S$~\ref{sec:supernova-rate} and
$\S$~\ref{sec:supern-prop-as}. In $\S$~\ref{sec:supernova-rate} we
examine the rate of SNe~Ia as a function of various host galaxy
parameters, and in $\S$~\ref{sec:supern-prop-as} examine the impact of
environment on SN~Ia light-curve shape parameters.  We summarize in
$\S$~\ref{sec:conclusions}.

Throughout the paper, we assume a cosmology of $\omatter=0.3$,
$\olambda=0.7$,
$\mathrm{H}_0=70\,\mathrm{km}\,\mathrm{s}^{-1}\,\mathrm{Mpc}^{-1}$,
and use the AB photometric system \citep{1983ApJ...266..713O}.

\section{The SNLS dataset}
\label{sec:snls-dataset}

This section describes the observational dataset that we use for our
analysis in this paper. We first describe the dataset of SNe~Ia, and
then the datasets of both the SN~Ia host galaxies, and the general
galaxy population.

\subsection{Supernova data}
\label{sec:supernova-data}

Our SN~Ia data come from the Supernova Legacy Survey (SNLS). The SNLS
is a ``rolling'' search for distant SNe with a primary science goal of
using 700 high-redshift SNe~Ia to determine the average
equation-of-state parameter of dark energy, $<w>$
\citep[see][]{2006A&A...447...31A}. SNLS exploits the square-degree
Megacam camera \citep{2003SPIE.4841...72B} on the Canada-France-Hawaii
Telescope (CFHT) to conduct repeat $g'r'i'z'$ imaging of 4 low
Galactic extinction fields (named D1 to D4, see
\citealt{2006AJ....131..960S} for the field coordinates), imaged as
part of the ``deep'' component of the five year CFHT Legacy Survey
(CFHTLS). The data are time-sequenced with observations conducted
every 3-4 nights in dark time, allowing the construction of high
quality multi-color SN light-curves.  Supplementary $u^*$-band data is
also acquired, though these data are not time-sequenced.  SN
candidates are observed spectroscopically at the ESO-VLT, Gemini or
W.~M.~Keck telescopes to confirm their nature and to measure a
spectroscopic redshift either from host galaxy features or from the SN
itself \citep[see][for an overview of the spectroscopic
typing]{2005ApJ...634.1190H}. A description of the real-time search
operations and the criteria for following SN candidates
spectroscopically can be found in \citet{2006AJ....131..960S}.

The survey is now in its third year, and each of the four survey
fields has now been observed for at least two complete ``seasons'',
each spanning the 5-6 months the field is visible from Mauna Kea.
This paper uses data from the first two seasons of D1, D2 and D4, as
well as the second and third seasons of D3. (The first season of D3
was performed in a pre-survey period where the data quality and
spectroscopic completeness were substantially lower.) During this
period, 116 spectroscopically confirmed SNe~Ia were obtained over the
redshift range $0.2<z<0.75$, of which 100 were detected during the
period of the survey for which an accurate rate efficiency calculation
can be performed \citep[][and
$\S$~\ref{sec:sn-incompl-corr}]{2006neill}.  To this confirmed SN~Ia
sample, we add 24 probable SNe~Ia (see $\S$~\ref{sec:sn-incompl-corr})
which lack a spectroscopic identification but for which a reliable
SN~Ia photometric redshift can be estimated and which possess an
excellent light-curve fit to a SN~Ia. This set of 124 SNe~Ia forms the
primary sample studied here.

\subsection{Galaxy data}
\label{sec:host-galaxy-data}

The images that we use to measure the galaxy parameters are also
constructed from data taken as part of the CFHT-LS. Deep optical
stacks were generated for each filter ($u^*g'r'i'z'$) in each field
from ``Elixir'' processed images \citep{2004PASP..116..449M} available
from the Canadian Astronomy Data Centre
(CADC)\footnote{http://cadcwww.dao.nrc.ca/}. We summarise the main
details of our processing steps here. A precise astrometric solution,
accounting for distortion, is assigned to every image frame, and the
seeing and photometric quality determined by flux measurements of
tertiary standard stars in the CFHT-LS fields compiled by the SNLS
team.  Any two-dimensional sky variation is removed from each frame by
fitting the background spatially and subtracting the resultant fit.
Each individual frame has a weight-map associated with it, containing
the uncertainty in each pixel from considerations of photon noise from
the sky background and object photons. Known bad pixels (as determined
from the Elixir flat-fields) and saturated pixels are assigned a
weight of zero.

For the SN host galaxies, we require deep optical stacks with no SN
light present when we measure the fluxes of a given object (the
presence of SN light would otherwise bias the galaxy flux
measurements). We therefore generate stacks on a ``per-season'' basis;
stacks for season one SNe are generated using only data taken in
season two and later seasons, and so on.  The period between observing
seasons for each field -- 6 to 7 months -- allows ample time for SN
light to have faded to an negligible level.  Every image is resampled
to a common pixel co-ordinate system, excluding images that do not
meet the data quality criteria for a given field/filter/season
combination. The goal is to maximize exposure depth whilst retaining
excellent seeing in the final stacks (hence different seeing cuts were
used for each field/filter/season to account for varying observing
conditions at different times of each year).  We also require that for
a given field/season combination, the stacks in each filter have a
similar seeing. The typical seeing of the final stacks is
0.7-0.8\arcsec. The SWARP package (version 2.15.6) with a LANCZOS3
kernel is used for the resampling. The resulting resampled frames are
combined using a weighted average, with a sigma-clipping to remove
artifacts (from satellite trails and cosmic-rays) using a
custom-written routine.

We also require flux information for the general galaxy population to
act as a comparison sample to the SN~Ia host galaxies in later
sections. For this we use data from the two fields for which the
deepest data exists (D1 and D4), particularly in $u^*$. We use the
same stacking algorithm as above, but produce stacks including all the
data for a given field -- the field galaxies have no contaminating SN
light to concern us -- hence these stacks are deeper than those used
for the host galaxies.

Photometric zeropoints are derived from observations of standard stars
taken from \citet{2002AJ....123.2121S}, and transformed to the Sloan
Digital Sky Survey (SDSS) photometric system; these zeropoints are
determined separately for each of the deep fields in the month in
which the photometric reference epoch is defined. Two of the fields
(D2 and D3) are in the SDSS, and this provides an independent check of
the zeropoints. The CFHT-LS observes in a filter system close to the
\citet{2002AJ....123.2121S} $u'g'r'i'z'$ SDSS system, though the
Megacam $u^*$ filter differs from the SDSS $u'$ filter, being designed
to take advantage of the improved UV capabilities of CFHT and Megacam.
Our effective filter responses can be found in
\citet{2006AJ....131..960S}.

\subsection{Host galaxy identification and flux measurement}
\label{sec:host-galaxy-ident}

The correct host galaxy for every SN~Ia in our sample is identified as
follows. We begin by measuring the SN position from point-spread
function (PSF) fits to the SN in subtraction images on epochs when the
signal-to-noise is largest. These SN positions are accurate to a
fraction of a pixel, or $\simeq0.02$\arcsec\ (one megacam pixel is
0.186\arcsec). By converting these pixel positions to RA and DEC, we
can then precisely identify the location of the SN on the
$u^*g'r'i'z'$ deep stacks even though no SN light is present, as all
the frames share a common astrometric system. For each SN~Ia, we use
SExtractor 2.4.4 \citep[][see \citealt{2005astro.ph.12139H} for an
excellent user guide]{1996A&AS..117..393B} to identify all potential
host galaxies within 10\arcsec\ of the SN position, and simultaneously
measure photometric information on the host galaxies.  SExtractor is
used in dual-image mode for the photometric measurements; detections
are performed in the $i$' filter (the filter with the deepest data)
and measurements performed in each of $u^*g'r'i'z'$.

Various photometric measurements and galaxy structural parameters are
recorded for each galaxy. We use the SExtractor MAG\_AUTO flux
measures and associated errors throughout this paper. The MAG\_AUTO
aperture is a flexible elliptical aperture
\citep[e.g.][]{1980ApJS...43..305K} with a characteristic ``Kron''
radius; we configure SExtractor to measure fluxes inside 2.5 Kron
radii.  Analytically, this results in $\sim$90\% of an object's light
being measured in each aperture
\citep{1987A&A...183..177I,2005PASA...22..118G}, though this may break
down for very faint, highly concentrated galaxies at the detection
limit \citep[][see discussion in
\citealt{2005PASA...22..118G}]{2002ApJ...571...56B,2004ApJS..150....1B}.
The use of SExtractor in dual-image mode (and the similar PSFs of the
stacks in the different filters) ensures that the same size MAG\_AUTO
aperture is used for the different measurements of a given galaxy,
minimizing aperture mismatches. Statistical errors in each flux
measurement are estimated using the weight-image of the final stack as
an r.m.s image in SExtractor.

\begin{figure}
\plotone{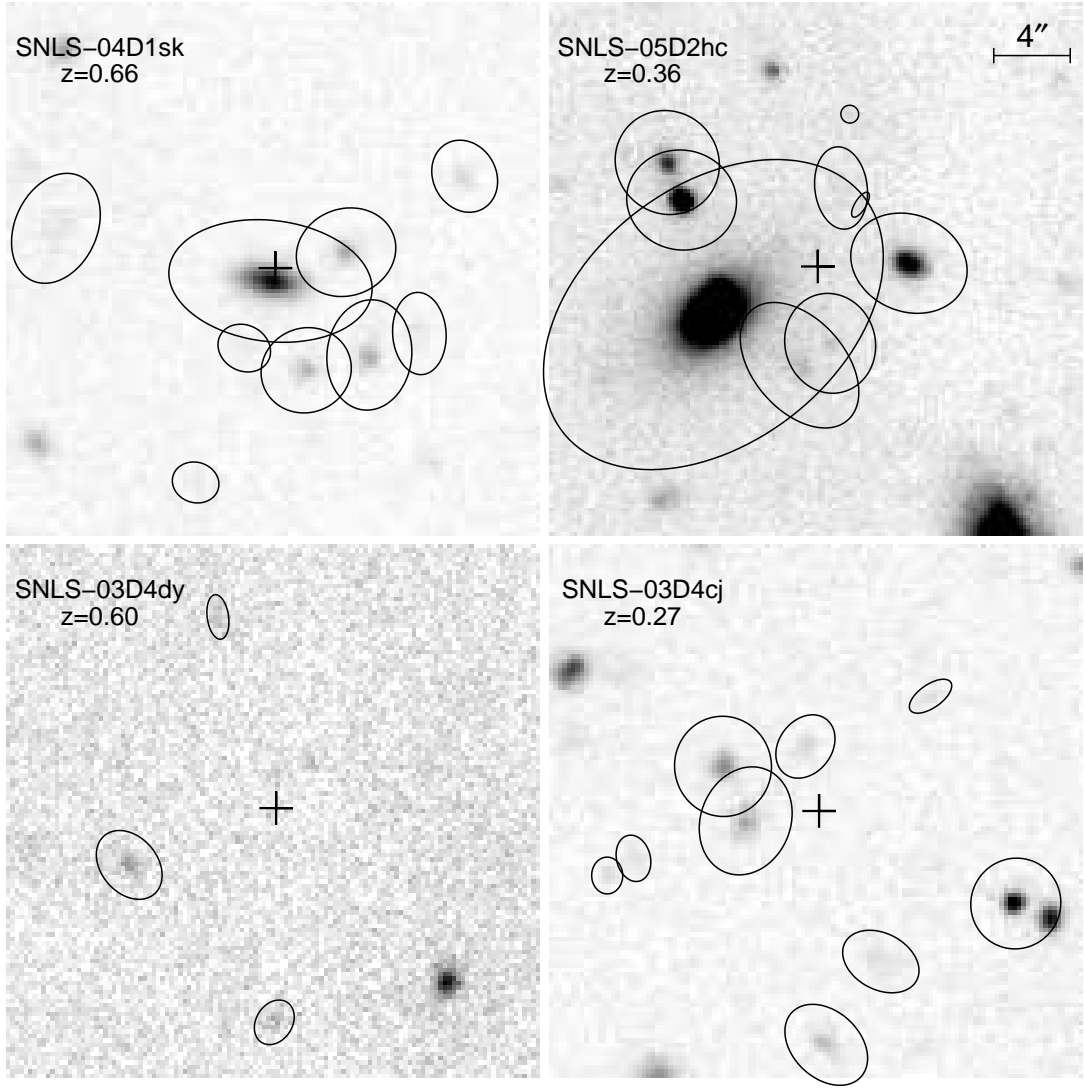}
\caption{
  Examples of the SN~Ia host galaxy identification technique used in
  this paper (see $\S$~\ref{sec:host-galaxy-ident}). Four SNe are
  shown. In each panel, the SN position is marked with a cross, and
  each candidate host, as detected by SExtractor, has the $5R$ ellipse
  over-plotted (see $\S$~\ref{sec:host-galaxy-ident} for the
  definition of $R$). The nearest host in terms of this $R$ parameter
  is considered to be the correct host; SNe with no hosts inside $5R$
  are considered ``hostless''. Top Left: SNLS-04D1sk, a
  straightforward case where the identification is unambiguous. Top
  Right: SNLS-05D2hc, a case where the nearest host in terms of
  arcseconds is probably not the correct identification. Bottom Left:
  SNLS-03D4dy, a case where no potential host is found within several
  arcseconds of the SN position.  Bottom Right: SNLS-05D2hc, where all
  candidate hosts lie at $R>5$.
\label{fig:montage}
}
\end{figure}

The identification of the correct host galaxy is not always
straightforward -- occasionally the closest host in arcseconds is
probably not the correct identification.  Instead we use a separation
normalized by the apparent size of the galaxy to which the SN is being
compared (Figure~\ref{fig:montage}). We calculate the separation of the SN from each candidate
host galaxy in terms of the elliptical radius ($R)$ along a line
connecting the SN pixel position ($x_{\mathrm{sn}}$,$y_{\mathrm{sn}}$)
to a given host center ($x_{\mathrm{gal}}$,$y_{\mathrm{gal}}$).  The
elliptical shape is determined by SExtractor, defined by semi-major
($r_A$) and semi-minor ($r_B$) axes together with a position angle
($\theta$), with $R$ given by

\begin{equation}
R^2={C_{xx}}x_r^2 + {C_{yy}}y_r^2 + {C_{xy}}x_ry_r,
\end{equation}

\noindent
where $x_r=x_{\mathrm{sn}}-x_{\mathrm{gal}}$,
$y_r=y_{\mathrm{sn}}-y_{\mathrm{gal}}$,
$C_{xx}=\cos^2(\theta)/r_A^2+\sin^2(\theta)/r_B^2$,
$C_{yy}=\sin^2(\theta)/r_A^2+\cos^2(\theta)/r_B^2$, and
$C_{xy}=2\cos(\theta)\sin(\theta)(1/r_A^2+1/r_B^2)$. $R$ is determined
for every candidate host galaxy, the hosts ordered by $R$, and each SN
is assigned to the host that is nearest in terms of this parameter.
The isophotal limit of a given object corresponds to $R\sim3$; we
consider only galaxies with $R\leq5$. In some cases ($\sim$7\%), no
host galaxy is identified by SExtractor with $R\leq5$; we then measure
the flux and flux error inside a 3\arcsec\ diameter aperture centered
on the SN position, and use this as the ``galaxy'' flux in what
follows. Figure~\ref{fig:montage} shows a visualization of the
technique for four SNe~Ia.

We also measure the properties of the general galaxy population using
the deeper stacks constructed as described above. We use SExtractor
with the same parameters as for the SN host galaxies.

\section{Galaxy Properties}
\label{sec:galaxy-meas}

This section details the techniques we use for converting the galaxy
properties into characteristics that can be used for our subsequent
analysis in $\S$~\ref{sec:supernova-rate} and
$\S$~\ref{sec:supern-prop-as}. We first discuss the technique of
converting the observed fluxes of $\S$~\ref{sec:host-galaxy-ident}
into galaxy stellar masses and star-formation rates, both for the host
galaxies themselves and for the general field galaxy population. We
then detail a number of consistency checks we perform on these derived
parameters.

\subsection{Galaxy SED fitting}
\label{sec:galaxy-sed-fitting}

In order to investigate SNe~Ia in relation to the mean properties of
the stellar population from which they were formed, we need to derive
properties such as stellar mass and star-formation rate (SFR) for the
galaxies. To do this we fit a series of template galaxy spectral
energy distribution (SEDs) to the broadband fluxes available for each
galaxy ($\S$~\ref{sec:host-galaxy-ident}), and then use the best-fit
SED for the estimation of the various properties \citep[see for
example][]{2000ApJ...536L..77B}. This technique is similar to that
employed by photometric redshift codes
\citep[e.g.][]{1996ApJ...468L..77G,2000A&A...363..476B,2002A&A...386..446L}.
The best-fit SED is determined using a $\chi^2$ minimization between
the observed fluxes, the corresponding flux errors, and the synthetic
photometry generated by integrating the template SEDs through the SNLS
effective filter responses. For the host galaxies, the spectroscopic
redshift is known from the SN confirmation spectrum and held fixed,
reducing the uncertainty in the derived properties of each galaxy. For
the field population no spectroscopic information is available, and
the redshift is left as a free parameter in the fits.

Our set of synthetic templates is computed with the P\'EGASE.2 galaxy
spectral evolution code
\citep{1997A&A...326..950F,2002A&A...386..446L,2004A&A...425..881L}.
This code and the SEDs that it generates have been used extensively in
the literature to constrain the properties of high-redshift galaxies
\citep[e.g.][]{2004Natur.430..181G,2004ApJ...614L...9M,2006astro.ph..3094G}.
We use eight scenarios which evolve self-consistently with age;
details of the physical parameters defining them can be found in Table
1 of \citet{2002A&A...386..446L}.  We assume a
\citet{2001MNRAS.322..231K} initial mass function (IMF), and assume
the IMF is universal across environment \citep[see discussion in
][]{2002Sci...295...82K}. The synthetic SEDs are the sum of light
emitted by stars and nebular emission (continuum and lines), including
attenuation by dust with a King model or a plane-parallel slab
geometry. The quantity of dust evolves consistently with the amount of
gas present in the galaxy.

An SED is computed at 69 time-steps in each of the 8 scenarios, giving
a total of 552 template SEDs. When fitting a given galaxy, we use only
templates younger than the age of the Universe at the redshift of the
galaxy.

\subsection{Galaxy derived properties}

Two physical parameters, for both the SN~Ia host galaxies and for the
general galaxy population, are of particular interest in this study:
the galaxy mass, and the galaxy SFR. The first is the total stellar
mass of a galaxy, the total current mass in stars of all types and
ages. This is derived by integrating the total star-formation history
(SFH) of the best-fit scenario up to the best-fit age, and subtracting
the mass of stars that have died.

The second parameter is the amount of recent star-formation that the
galaxy has experienced.  There are several approaches that could be
taken. First, one could simply divide the ultra-violet (UV) luminosity
of each galaxy by a conversion factor to obtain a UV-derived SFR
\citep[e.g.][]{1987A&A...180...12D}. However, such an approach tends
to over-estimate the SFR in old stellar systems due to the
contaminating presence of evolved stars.  A second approach to
determining instantaneous SFRs is to use nebular emission lines
\citep{1998ARA&A..36..189K}.  Unfortunately, only very limited
information on the appropriate lines is available for our host sample
from the SN confirmation spectra, with large uncertainties from
aperture losses, SN light contamination and flux calibration. 

We instead estimate the \textit{mean} SFR from our best-fitting
scenarios, averaging the SFR over a longer time interval.  This
provides an automatic correction for the UV light from old evolved
stars. To define the period over which the star-formation should be
averaged, we carried out simulations where the mean SFR was estimated
from synthetic $u^*g'r'i'z'$ photometry measured both on the idealized
models and on models with stochastic SFHs (see
$\S$~\ref{sec:systematic-errors}). These simulations showed that mean
SFRs on periods of 0.5\,Gyr can be reliably recovered without
significant systematic errors. On shorter timescales systematics can
be introduced, particularly for galaxies for which the redshift is not
known (see Table~\ref{tab:simulations} and discussion in
$\S$~\ref{sec:systematic-errors}).

For the general galaxy population, an alternative approach would be to
use a parameterized form of the mass or luminosity functions, either
directly inferred from high-redshift spectroscopic surveys
\citep[e.g.,][]{2005A&A...439..863I,2005astro.ph..6041W,2005ApJ...625..621B}
or low-redshift functions adjusted for evolution to high redshift.
While this is possible for the global mass function, no equivalent
distribution exists for the SFR property of field galaxies, requiring
a conversion from (for example) a rest-frame $U$-band or UV luminosity
function.  Furthermore, since a goal of this work is to investigate
host galaxy properties binned by the type of the host galaxy
($\S$~\ref{sec:supernova-rate}), mass and luminosity functions derived
from galaxies categorized in the same way would be required, and these
are not currently available.

Uncertainties in the derived parameters for our galaxies arise from
both statistical and systematic errors (discussed in the next
section). Statistical errors derive from the photometric measurements
of the galaxies, and are accounted for by considering the full range
in the quality of the 552 template fits (as defined by the $\chi^2$
statistic). This maps out a probability surface in the stellar mass
and SFR parameters determined from the fits, and allows estimates of
the statistical uncertainties of each of the derived physical
parameters; lower signal-to-noise photometry will be consistent with a
wider range of fit templates and hence a wider range of stellar masses
and SFRs.  Invariably, the derived masses and SFRs possess
non-symmetrical error-bars, particularly for the fainter systems. In
some cases, fainter galaxies can have more than one minimum in the
$\chi^2$ space.  In this case, the best-fit is considered as the
solution, but the error bars in mass and SFR are extended to the
largest interval of uncertainty, covering the range of parameters
between the minima.  The errors are therefore conservative for
galaxies with several minima.

\subsection{Systematic errors}
\label{sec:systematic-errors}

Systematic errors in our derived galaxy stellar masses and mean SFRs
potentially arise from many sources. These include the wavelength
range of our input photometry, and our choice of spectral libraries,
IMFs and SFHs that form the template SEDs used for the fitting, and
are of course harder to estimate. We first test the accuracy of the
photometric redshift estimates by running fits to the host galaxies
(with known spectroscopic redshifts) with the redshift left as a free
parameter.  Figure~\ref{fig:speczphotoz} (left) shows these
spectroscopic and photometric redshift estimates for the SN~Ia host
galaxies. For galaxies with SExtractor detections, the median of
$|\deltaz|=|\zspec-\zphot|$ is 0.02 (the 90\%-ile is 0.15), and the
median of $|\deltaz|/(1+\zspec)$ is 0.012 (90\%-ile is 0.09). These
measures contain photometric redshifts from galaxies with a
significant photometric uncertainty due to their faintness (and hence
uncertainty in the photometric redshifts). To confirm that these error
estimates are reasonable, we calculate the dispersion
$|\deltaz|/\sigma$, where $\sigma$ is the appropriate error in the
photometric redshift. 90\% of the photometric redshift estimates lie
within 2$\sigma$ of their corresponding spectroscopic redshift.

We next compare the host galaxy properties derived from the template
fit with a known redshift against those derived from the photometric
redshift fit \citep[c.f. figure~3 of][]{2005ApJ...625..621B}. Such a
comparison will show whether the photometric-redshift properties of
the galaxies we study can reliably trace the true galaxy properties in
the mean. We compare the stellar mass and mean SFR estimates for
galaxies using the spectroscopic redshift with the same estimates when
the redshift is allowed to float
(Figure~\ref{fig:speczphotoz_masscompare}). The agreement for both
quantities is reasonable, with the mass estimates appearing more
robust. No mass or SFR-dependent trends are seen.  The median
difference $\Delta M=M_{\mathrm{spec}}-M_{\mathrm{phot}}$ is 0.026 dex
(90\% of $\Delta M$ lie within 0.2 dex), and the difference $\Delta
\mathrm{SFR}$ is 0.033 dex (90\% of $\Delta \mathrm{SFR}$ lie within
0.38 dex). A comparison of the mass function derived from our
photometric redshift fits to published mass functions of field
galaxies will be presented in an upcoming paper (Le Borgne et al., in
prep).

The next potential source of concern is the lack of near-IR data and
the impact on our estimates of stellar masses, and the lack of very
short wavelength data and the impact on the derived SFRs. These could
be particularly pronounced in systems with both a young and an old
stellar populations, such as those galaxies experiencing recent
star-formation events. Clearly the uncertainties in derived stellar
masses and mean SFRs will be larger in the absence of UV and near-IR
data; however as long as the error-bars in these properties reflect
this then this additional uncertainty will be carried through in our
analysis. A larger concern would be any systematic under- or
over-estimation of the stellar masses and SFRs.

We investigate this by using our standard templates to fit photometry
generated from SEDs with stochastic SFHs. Synthetic SEDs are formed
from the composite of 3 random exponential SFHs each with a different
age ($t$), mass fraction and $\tau$. One SFH is constrained to be old
($t$$>$1Gyr), one of intermediate age (200Myr$<$$t$$<$2Gyr) and one
young (30Myr$<$$t$$<$100Myr). Each of these 3 random SFHs is converted
into an SED at the randomly-selected $t$, and the 3 SEDs added
together to form one composite SED. We place this composite SED at a
random redshift (0.2$<$$z$$<$0.75), generate synthetic $u^*g'r'i'z'$
photometry, and see how well P\'EGASE.2 recovers the stellar mass and
mean SFR using the 8 idealized scenarios of
$\S$~\ref{sec:galaxy-sed-fitting} both when the redshift is fixed and
when it is left as a free parameter.

The results are given in Table~\ref{tab:simulations} and are
encouraging. For the stellar masses, we see only small differences in
the mean between the input and recovered value with a standard
deviation of around 0.3dex. For the mean SFRs, again there is no
significant systematic offset though the standard deviation of the
differences is larger particularly when the redshift of the galaxy is
not known.

We also briefly investigate the effect of dust on our stellar mass and
mean SFR estimates. Though dust is included in the idealized scenarios
(see $\S$~\ref{sec:galaxy-sed-fitting} and discussion in
\citealt{2002A&A...386..446L}), this will only provide an average
correction to our derived properties. We experimented with adding
\textit{extra} dust to our simulated SEDs (up to E(B-V)=0.5 using a
\citealt{2000ApJ...533..682C} extinction law). As expected, this can
lead to a systematic under-estimation of the mean SFR (up to 0.4 in
dex in some cases) and a small increase in the standard deviation,
though the accuracy of the stellar mass estimates was not affected.

Finally, we estimate systematic uncertainties by deriving the galaxy
properties using a similar technique but different templates. We use
the photometric redshift code of \citet{1996ApJ...468L..77G}, which
uses a different set of (empirical) galaxy templates
\citep{1980ApJS...43..393C,1996ApJ...467...38K} to those used in
Z-PEG. We determine masses by fitting \citet{2005MNRAS.361..725B}
population models to the best-fitting templates which results in a
mass-to-light ratio which can be used to convert the galaxy
luminosities into masses. SFRs are determined by dividing the flux at
2800\AA\ by $4.8\times 10^{27}$ \ergshz\ 
\citep*{2002AJ....123.1188B,1998ApJ...498..106M}, with a small
correction for early-type SEDs to account for the contribution from
older stars \citep[see][]{2002AJ....123.1188B}

The two different techniques for estimating the masses and the SFRs
showed a good agreement. The mass and SFR estimates agree to better
than a factor of two -- mean difference in mass was 0.35 dex (in the
sense Z-PEG measured smaller masses) with an r.m.s.  scatter of 0.24,
and the mean offset for the SFR estimates was 0.15 dex (in the sense
Z-PEG measured larger SFRs) with an r.m.s.  scatter of 0.29.  Most
importantly, no mass-dependent or SFR-dependent trends were detected.
Given that Z-PEG is used to calculate masses and SFRs for both field
and host galaxies, these offsets in mass or SFR will cancel in our
analyses.

\section{Incompleteness corrections}
\label{sec:incompl-corr}

In the next two sub-sections we discuss incompleteness in our sample.
We first discuss the incompleteness of our field galaxy population due
to the limiting magnitude of our deep field stacks, an effect common
to all galaxy redshift surveys. We then discuss the incompleteness of
our SN~Ia sample.

\subsection{Galaxy incompleteness corrections}
\label{sec:galaxy-incompl-corr}

Incompleteness will affect our stellar mass and mean SFR distribution
functions of the field galaxy population.  Our survey is magnitude
limited; galaxies with a given absolute magnitude (and spectral type)
will become fainter than this magnitude limit at different redshifts,
which may be less than the largest redshift we consider here
($z=0.75$). We adopt the traditional \vmax\ method
\citep[e.g.][]{1968ApJ...151..393S,1976ApJ...207..700F} to correct for
this effect.  \vmax\ is defined as the co-moving volume within which
each galaxy -- as defined by its absolute magnitude and $k$-correction
from best-fitting SED template -- would remain in our sample i.e.
within which it would satisfy the limits in apparent magnitude of the
current depths of the optical stacks. Each galaxy is then weighted by
a factor $V_{\mathrm{survey}}$/\vmax\ when computing the various mass
and SFR distributions used in $\S$~\ref{sec:supernova-rate}, where
$V_{\mathrm{survey}}$ is the total survey volume.

The SExtractor photometric measurements are performed
by detecting in the $i'$ filter, and the depth in this filter defines
the limiting magnitudes. The limiting magnitudes on the $i'$ D1 and D4
stacks are determined by inserting fake sources into the stacks, and
measuring the fraction recovered by SExtractor as a function of
magnitude; the limiting magnitude is that at which 50\% of fake
sources are recovered.

\subsection{SN incompleteness and rate calculation}
\label{sec:sn-incompl-corr}

Incompleteness will affect our SN~Ia sample in several ways. Firstly,
we could miss SNe~Ia altogether due to inefficiencies in our search
pipeline (the detection efficiency). Secondly, detected SNe~Ia in the
redshift range of interest could be lacking a spectroscopic
observation because such an observation would be too challenging, for
example due to host galaxy contamination or the SN maximum light
falling during a bright-moon period. Finally, SNe~Ia could lack a
spectroscopic confirmation because poor weather affected the
spectroscopic scheduling.

To correct for these effects, we adopt the scheme of
\citet{2006neill}, who calculated the SNLS SN~Ia rate over
$0.2<z<0.6$, but we extend the calculation to $z=0.75$. The first
source of incompleteness, detectability, is accounted for using
simulated SN data. Fake SNe are placed in real SNLS data, and the
recovery efficiency as a function of SN brightness, position, sky
background, host brightness, exposure time, seeing, and sky
transparency is calculated \citep[see][]{2006neill}.

We also determine the observing window within which maximum-light of a
SN~Ia would have to occur in order to be considered for a
spectroscopic observation and for which we can derive a reliable
photometric classification (see below). SN~Ia candidates detected at
the start or end of an observing season lack a full light-curve and
therefore are not usually observed spectroscopically as any resulting
light-curve fit would be far more uncertain. For the purposes of this
study, these cut-offs are defined as follows. i) The SN must have at
least 2 $i'$ and at least 1 $r'$ observation between -15 and
-1.5\,days in the SN rest-frame, ii) There must be at least 1 $g'$
observation between -15 and +5\,days, and iii) There must be at least
1 $r'$ or $i'$ observation after +11.5\,days but before +35\,days.
These criteria cull SNe~Ia whose light-curves do not properly sample
maximum light and for which a light-curve width measurement would be
correspondingly more uncertain, and for which a color near maximum was
not measured.  Any confirmed SNe~Ia which do not meet these criteria
are excluded from this analysis; this reduces our sample from 116
SNe~Ia to 100.  Our simulations then provide, for a given field
observing season, the efficiency required to convert an observed ``per
season'' rate (the number of SNe~Ia with 0.2$<$$z$$<$0.75 meeting the
light-curve criteria defined above) into a yearly rate. These
efficiencies can be found in Table~\ref{tab:eff_info}.

\begin{figure}
\plotone{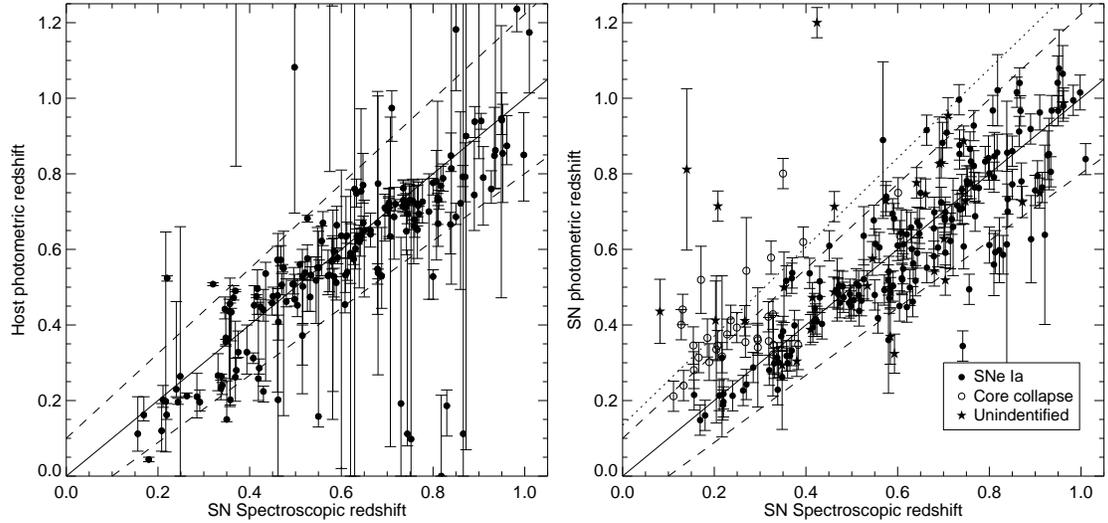}
\caption{
  The comparison of spectroscopic and host galaxy photometric redshift
  estimates for the SNLS SN~Ia host galaxies (left) and the comparison
  between spectroscopic and SN photometric redshift estimates for all
  SNLS confirmed SNe (right).  The solid line shows a 1:1 agreement
  between \zspec\ and \zphot, the dashed lines show a 10\% uncertainty
  in (1+\zspec). The right-hand plot, the dotted line shows the region
  above which SNe were excluded as SNe~Ia during the SN completeness
  correction stage ($\S$~\ref{sec:sn-incompl-corr}). The SN~Ia
  photometric redshifts follow the method of
  \citet{2006AJ....131..960S}.  For the host galaxy photometric
  redshift, the median of $\zspec-\zphot$ is 0.02 with a standard
  deviation of 0.15. For the SN photometric redshifts, these values
  are -0.007 and 0.088. Note that the SN~Ia photometric redshift
  estimates include a cosmological prior that precludes their use for
  determining the cosmological parameters.
\label{fig:speczphotoz}
}
\end{figure}

\begin{figure}
\plotone{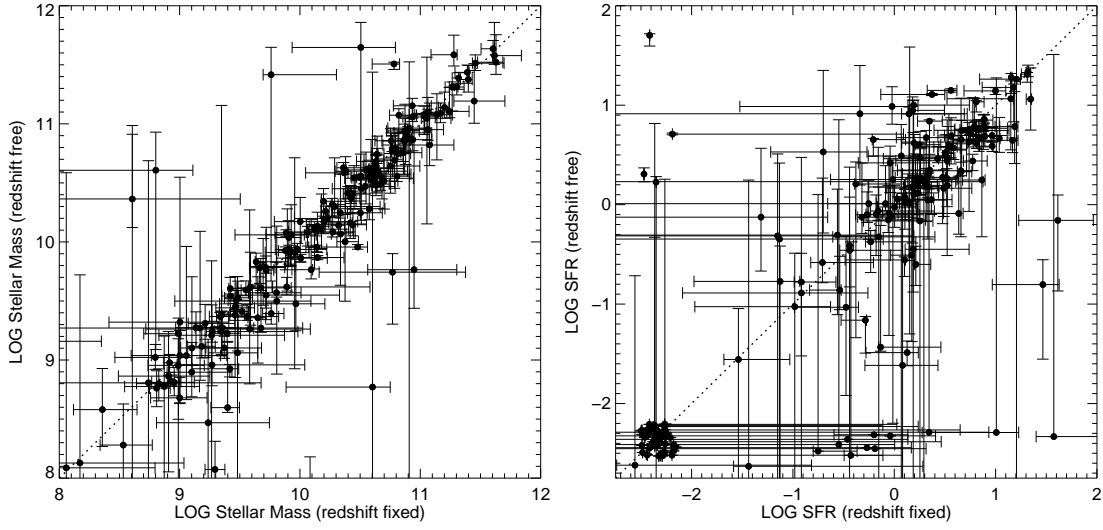}
\caption{
  The comparison of mass ($M$) and star-formation rate (SFR) estimates
  for the SNLS SN~Ia host galaxies when the redshift is fixed
  (spectroscopic redshift; y-axes), and when the redshift is
  determined using the photometric redshift code, Z-PEG (photometric
  redshift; x-axes). Zero SFR systems are shown with
  $\log(\mathrm{SFR})=-2.5$ plus a random offset for clarity. The
  average properties of the sample as determined when the redshift is
  known versus the photometric redshift estimate are very similar; the
  median difference $\Delta M$ is 0.026 dex (90\% of $\Delta M$ lie
  within 0.2 dex), and the difference $\Delta \mathrm{SFR}$ is 0.034
  dex (90\% of $\Delta \mathrm{SFR}$ lie within 0.38 dex).
\label{fig:speczphotoz_masscompare}
}
\end{figure}

The spectroscopic incompleteness -- the fraction of candidate SNe~Ia
detected but never observed spectroscopically -- is the most
challenging source of uncertainty to address in this study. We assess
the incompleteness using the photometric SN selection method presented
in \citet{2006AJ....131..960S}.  This technique fits SN light-curves
in the absence of a spectroscopic redshift, and returns the
best-fitting SN~Ia parameters (redshift, stretch, \ebmvhost, and a
dispersion in the SN peak magnitude, $\mathrm{d}m$), as well as a
guide that the candidate under study is a SN~Ia.  The software was run
on $g'r'i'z'$ photometry for all SN candidates (including those
observed and typed spectroscopically) discovered during the period
covered by the simulations above. A comparison of SN spectroscopic
redshift and SN photometric redshift for all SNLS SNe~Ia can be found
in Figure~\ref{fig:speczphotoz}. (Note that as the SN photometric
redshift code makes an assumption about the cosmology when fitting the
light-curve, the resulting photometric redshift cannot themselves be
used for determination of the cosmological parameters.)  The mean
difference $\zspec-\zphot$ for all SNe~Ia is -0.007; the standard
deviation is 0.088. For the SNe~Ia included in this study -- over
0.2$<$$z$$<$0.75 and which meet the light-curve criteria -- the
standard deviation is 0.077. Occasionally, a spectroscopic observation
of a SN candidate yielded a spectroscopic redshift but no definitive
type; in these cases we still allow the redshift to float for
comparison with the spectroscopic redshift.

There were 286 SN candidates without a spectroscopic confirmation with
a SN~Ia photometric redshift in the range of interest. 80 do not meet
the light-curve coverage criteria that we apply to the
spectroscopically confirmed sample, and are excluded. This ensures
that all candidates have a reliable stretch and $g'$ observations. We
also exclude 2 SN candidates where the spectroscopic redshift was in
disagreement with the photometric redshift where
$\zphot>\zspec+0.135\times(1+\zspec)$ (Figure~\ref{fig:speczphotoz}). In
these cases, the SN spectra were ambiguous and no type could be
determined; usually this was because a SN~Ib/c spectrum provided a
similar quality fit to the observed spectrum as did a SN~Ia.

\begin{figure}
\plotone{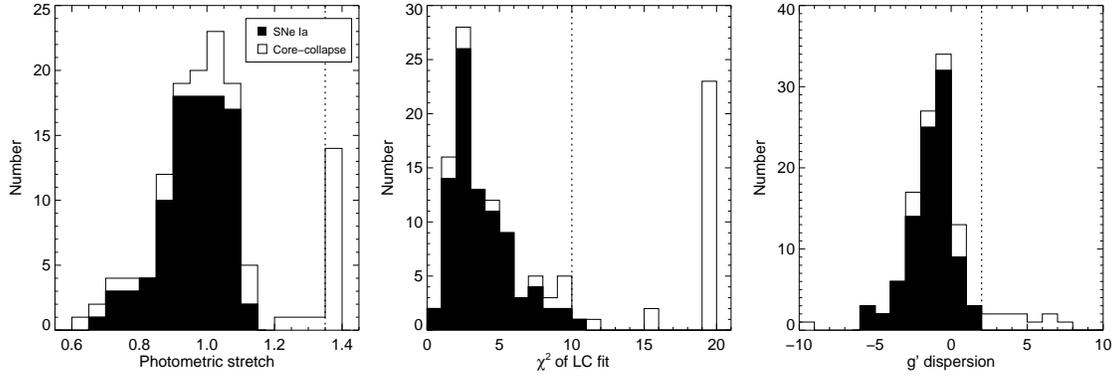}
\caption{
  The range of fitted stretch (left), light-curve $\chi^2$ per degree
  of freedom (center), and early-time $g'$ dispersion (right) obtained
  when running the photometric redshift code of
  \citet{2006AJ....131..960S} on all spectroscopically confirmed SNLS
  SNe~Ia (filled histogram) and SNLS core-collapse SNe (open
  histogram) over $0.2<z<0.75$. Cuts of $s>1.35$, $\chi^2>10$, and
  $g'$ mean dispersion $>$2 (see equation~(\ref{eq:1})) are effective
  at removing core-collapse SNe while retaining SNe~Ia; these cuts are
  shown as vertical dotted lines in the figure.
\label{fig:phottypes}
}
\end{figure}

The remainder (204) were then culled based on their light-curve
parameterization and the $\chi^2$ of the light-curve fits in the
different observed filters. We exclude 70 SN candidates with a fitted
stretch $>$1.35 (very effective at removing SNe~IIP), and 75
candidates whose $\chi^2$ per degree of freedom in the fit was $>$10.
None of our spectroscopically confirmed SNe~Ia had fit parameters in
either of these ranges (see Figure~\ref{fig:phottypes}).

These simple cuts leave 67 candidate SNe~Ia. A further visual
inspection of the remaining candidates revealed that although many of
these had acceptable overall $\chi^2$ fits, they were too blue in
$g'-r'$ before +5\,d when compared to a SN~Ia template.  Hence, our
final statistical cut removes 35 objects with a poor $g'$ fit at
early-times (-15$<$d$<$+5), or with a $g'$ mean dispersion

\begin{equation}
\label{eq:1}
\frac{1}{N_g}\sum^{N_g}_{i=1}\frac{(g^i_{obs}-g^i_{model})}{g^i_{err}}>2,
\end{equation}

\noindent
where $N_g$ is the number of $g'$ light-curve fluxes $g_{obs}$ with
error $g_{err}$ over -15$<$d$<$+5, and $g_{model}$ is the SN~Ia
template model.  This cut does not remove SNe that are too red
(negative dispersions), which could be indicative of extinction, only
those that are too blue. The various culls that we use are summarized
in Table~\ref{tab:culls}. This leaves 24 SN~Ia candidates to add to
the 100 spectroscopically confirmed SNe~Ia sample; details can be
found in Table~\ref{tab:eff_info}. For these SNe, the redshift of the
SN~Ia is taken to be the SN~Ia photometric redshift where no
spectroscopic redshift was available. The impact on our results of
these photometrically classified SNe~Ia is discussed in
$\S$~\ref{sec:supernova-rate}.

\section{The Supernova Ia rate as a function of host galaxy properties}
\label{sec:supernova-rate}

In this section, we examine the SN~Ia rate as a function of the
properties of their host galaxies, and use these rates to place
constraints on the composition of the total SN~Ia rate. We first
classify our SNe~Ia into three sub-groups based upon the nature of
star-formation in their host galaxy. The first group comprises SNe~Ia
located in host galaxies with a zero mean SFR from the SED fits
(``passive'' galaxies). We then use the specific star-formation rate,
sSFR, defined as the star-formation rate per unit stellar mass
\citep[e.g.][]{1997ApJ...489..559G,2000ApJ...536L..77B,2004MNRAS.351.1151B},
to classify the star-forming SN~Ia host galaxies. With units of
$\mathrm{yr}^{-1}$, sSFR is essentially a measure of the inverse of
the formation time-scale for a given galaxy: high sSFR galaxies will
form the mass in their stellar populations on shorter times than low
sSFR galaxies. The second group of hosts, defined as $-12.0 \le \log
(\mathrm{sSFR}) \le -9.5$, have a small or moderate amount of
star-formation relative to their stellar mass, and therefore are
likely to possess a substantial evolved stellar component as well as
young stars. The third and final group, with $\log (\mathrm{sSFR})>
-9.5$, have a large amount of star-formation relative to their stellar
mass, and therefore stellar populations that have a large component
made up of young stars. Broadly speaking, the second group tends to
comprise normal star-forming galaxies such as the Milky Way, whereas
the third group tends to include vigorously star-forming and lower
mass dwarf galaxies. The division is illustrated in
Figure~\ref{fig:mass_sfr}, which shows the distribution of all the
SN~Ia host galaxies in the stellar-mass/SFR plane.

An attraction of the SED-fitting technique is that galaxies can be
classified according to their star-formation properties without regard
to morphology. This has some advantages; in morphologically-selected
samples of distant galaxies, a significant fraction of spheroidals
have been shown to possess both ``blue cores'' and weak O\,\textsc{ii}
emission lines in their spectra
\citep{2001MNRAS.322....1M,2005ApJ...633..174T}, interpreted as a
signature of young stars in these galaxies. There is also some
evidence for recent star-formation in some early-type galaxies from
the near-UV color-magnitude relation \citep{2005ApJ...619L.111Y}. The
advantage of the effective color-selection of SED-fitting is that
rather than assume all early-type galaxies consist purely of old
stars, the evolutionary model fitting places no \textit{a priori}
constraint on the type of SFH that can be fit to a given galaxy. Young
populations are perfectly possible in morphologically spheroidal
galaxies, and hence any young stellar population which has a
significant impact on a galaxy's colors will be reflected in the
best-fitting galaxy SFH.

We examine how the SN~Ia rate varies as a function of this specific
SFR in Figure~\ref{fig:rate_ssfr}. The field galaxies are binned
according to their value of specific SFR, and the total field-galaxy
stellar mass in each bin of specific SFR is calculated. We use the
\vmax\ technique of $\S$~\ref{sec:galaxy-incompl-corr} to
incompleteness correct this distribution. We then bin the SN~Ia host
galaxies by specific SFR in the same way and calculate the number of
SNe~Ia per unit stellar mass as a function of the specific SFR of
their host galaxies.  In Figure~\ref{fig:rate_ssfr}, the rate in
passive galaxies, where the SFR is zero, is shown as a hashed area
starting at $\log (\mathrm{sSFR})=-12$ (the height of this hashed area
represents the statistical uncertainty in the measurement).

An increase in the rate of SNe~Ia per unit stellar mass with
increasing specific SFR of the host galaxy is clear.  The difference
between the rate in passive galaxies and the most vigorous
star-forming systems is about a factor of 10; furthermore, the
increase in the rate is a fairly smooth function of the specific SFR.
The general trend of Figure~\ref{fig:rate_ssfr} can be compared to
that observed by \citet{2005A&A...433..807M} in the local universe
using a morphologically classified sample of local SNe~Ia host
galaxies. We illustrate this in Figure~\ref{fig:rate_ssfr} by
over-plotting the Mannucci et al. data on the SNLS results; the
Mannucci et al.  evolution is shown normalized to the SNLS rate in
passive galaxies.  The trend is very similar, though there is the
obvious caveat that the link between specific SFR and galaxy
morphology is not a straightforward one-to-one mapping.

As noted by other authors, this relationship is difficult to reconcile
with a model for SNe~Ia that originates solely from an old evolved
stellar population. \citet{2005astro.ph.10315M} and SB05 instead model
the SN~Ia rate (\snrIa) as a composition of two separate components: a
``prompt'' component, with a short delay time, and an ``old''
component, with a long delay time.  The most general form for the
\snrIa\ as a function of time is simply the convolution of the SFR,
${\dot M_{new}}$, and the probability function for getting a SN~Ia
from a stellar population of age $t$, $P$, i.e.

\begin{equation}
\label{eq:2}
\snrIa(t)=\int_0^t {\dot M_{new}}(t')P(t-t')dt'
\end{equation}

\noindent
This rate can be simply modeled by making the assumption that $P$ can
be well represented by two components. One has a peak of $B$ at time
$t=0$ and is zero at all other times (this represents a very short
delay time), the other has $P=A$ constant with time (and represents
long delay times), i.e.

\begin{equation}
\label{eq:3}
\snrIa(t)=A\int_0^t {\dot M_{new}}(t)dt + B{\dot M_{new}}(t).
\end{equation}

As $M_{\mathrm{tot}}(t)=\int_0^t {\dot M_{new}}(t)dt$, where
$M_{\mathrm{tot}}(t)$ is the total mass of a galaxy at time $t$, this
equation models the probability of a SN~Ia exploding in a given galaxy
as depending on both the mass and the instantaneous SFR of that
galaxy.  $A$ and $B$ are constants which relate the total mass and the
SFR of a galaxy to the \snrIa\ in that galaxy (which SB05 fix using
the observations of \citet{2005A&A...433..807M}). In effect, $A$ is
the \snrIa\ per unit mass of the old component, and $B$ the \snrIa\ 
per unit SFR of the young component. The model predicts that \snrIa\ 
is linear in both host galaxy mass and SFR.

Note that the definition of mass above is slightly different to that
measured for the SN host galaxies $\S$~\ref{sec:galaxy-sed-fitting}.
For the hosts, we measure the total \textit{stellar} mass i.e. the
total mass \textit{currently} in stars in each galaxy.
$M_{\mathrm{tot}}$ above is the integral of the SFH for each galaxy --
no correction is made for stars that have lost mass at the end of
their stellar evolution.  The numerical differences in these mass
definitions are shown in Figure~\ref{fig:mass-definitions}, with the
differences being largest in older, lower SFR systems.

The model of equation~(\ref{eq:3}) is a simplification of the real
physics.  For the prompt component, the model implies a zero
delay-time between star-formation and SN~Ia explosion; some non-zero
delay-time to account for main sequence lifetime and subsequent
accretion onto the white dwarf is obviously required. For the old
component, the equation simplifies the complex relationship between
the SN~Ia delay-time and the age of the stellar population by using a
simple constant probability rather than a more complex exponential or
Gaussian delay time distribution. In reality, for a coeval population,
after a few billion years the probability of a SN~Ia will likely
decrease as the stellar population ages and fewer progenitor stars are
available; this could cause an over-estimation of the SN~Ia rates in
the oldest, passive stellar systems.

Yet, these approximations may not be that poor.
\citet{2005astro.ph.10315M} find that the local SN~Ia delay-time
distribution is well represented by a prompt component modeled as a
Gaussian centered at $t=50$\,Myr plus a component modeled as an
exponential with a decay time of 3\,Gyr. These two terms can be
approximated by a delta function at time $t=0$ plus a constant
probability thereafter.  Furthermore, this equation parameterizes the
rate in a convenient form relating to galaxy properties that are
relatively straightforward to measure using the SNLS dataset
introduced above.

In the next sections, we illustrate this model by considering the
SN~Ia rate as a function of both host galaxy stellar mass and host
galaxy SFR taken separately, and then constrain the $A$ and $B$
parameters using a bivariate fit to the mass/SFR data.

\subsection{SN~Ia rate as a function of host galaxy stellar mass}
\label{sec:sn-rate-mass}

We first attempt to separate any component of the SN~Ia rate which may
depend on the stellar mass of a galaxy, from any component which
depends on the amount of recent star-formation i.e. to separate the
two components of equation~(\ref{eq:3}). We can do this using our
sample of SNe~Ia which exploded in passive galaxies, and which in our
models have zero recent star-formation activity. By binning the SN~Ia
host galaxies according to their mass, and dividing by the equivalent
(incompleteness corrected) distribution of the general galaxy
population in each mass bin, we can examine how the probability of a
SN~Ia explosion is related to the mass of the host galaxy.

We perform this comparison in Figure~\ref{fig:rate_mass}, where we
show the dependence of the SN~Ia rate as a function of mass of the
host galaxy. To account for the (invariably non-symmetrical) errors in
the mass determinations of the SN host galaxies, we have performed a
Monte-Carlo simulation with 5000 realizations of each SN host with the
masses for each host drawn from the estimated probability distribution
for that host. We then bin this Monte-Carlo population, normalize to
the total number of hosts in the sample, and use this binned
distribution in our analysis. Such a procedure produces a distribution
which accounts for observational statistical uncertainties, which can
be considerable in the fainter host sub-sample where some of the
measurements effectively provide only limits on the host parameters.

One feature of Figure~\ref{fig:rate_mass} is the increase in the SN~Ia
rate with increasing host galaxy mass, present among host galaxies of
all types.  As the SFR of the passive galaxies is zero (this defines
their selection), the contribution from the $B$ term in
equation~(\ref{eq:3}) is also zero in these galaxies. Hence the
best-fitting line in log-log space should have a slope of 1, if
equation~(\ref{eq:3}) is a good approximation -- the model is linear
in mass. For passive galaxies, we find that the slope,
$n_{\mathrm{mass}}$, is $n_{\mathrm{mass}}=1.10\pm0.12$ with a reduced
$\chi^2$ of $\simeq1.14$ (if only spectroscopically confirmed SNe~Ia
are used in the fits, $n_{\mathrm{mass}}=1.04\pm0.13$).  This implies
that the relationship between the SN~Ia rate and galaxy mass is
consistent with being linear and the model provides and adequate fit
to the data.  Using this technique we estimate
$A=5.1\pm1.2\times10^{-14}$\,\Aunits.

The relationship in later-type, star-forming galaxies is different.
The best-fitting slopes are statistically consistent: $0.66\pm0.08$
(reduced $\chi^2=0.77$) and $0.74\pm0.08$ (reduced $\chi^2=0.94$) for
the low and high specific SFR galaxies respectively. This translates
to an excess of SNe~Ia in low mass star-forming galaxies compared to
the passive galaxies, but a similar number in each at the most massive
end. This is as expected if the model of eqn.~(\ref{eq:3}) is correct.
Doubling the mass in passive systems will double the SN~Ia rate, but
doubling the mass in star-forming systems will only double the rate if
the SFR doubles as well. As the most massive star-forming systems
generally have lower specific SFRs than the lowest mass systems, the
increase in the SN~Ia rate with host mass in star-forming systems is
not linear. The result is that while the prompt component SNe dominate
in low-mass star-forming galaxies, the old component is more important
in higher-mass star-forming galaxies.

\subsection{SN~Ia rate as a function of host galaxy mean SFR}
\label{sec:sn-rate-sfr}

Having examined the SN~Ia rate in galaxies dominated by old stellar
populations, we now examine the SN~Ia rate in very young populations
via the SFR, which we determine as the mean SFR over the last 0.5\,Gyr
of a galaxy's SFH as determined from the best-fitting template
scenario ($\S$~\ref{sec:galaxy-sed-fitting}). We first bin both the
SN~Ia hosts and the general galaxy population by their mean SFR.  We
incompleteness correct the general galaxy population, and in each bin
of mean SFR for this general galaxy population, we sum the total
stellar mass in that bin, and, using a value of $A$ determined from
the passive galaxies in $\S$~\ref{sec:sn-rate-mass}, calculate the
number of SNe~Ia expected from any old component. This old component
is subtracted from each bin, leaving an excess of SNe~Ia above that
predicted from the stellar mass of the galaxies. We show this
distribution in Figure~\ref{fig:rate_sfr}. In this figure, we have
combined the SNe~Ia occurring in all types of star-forming galaxies
rather than the two separate star-forming populations shown in
Figure~\ref{fig:rate_mass}.  As in $\S$~\ref{sec:sn-rate-mass}, a
Monte-Carlo simulation is used to account for the uncertainties in the
stellar mass and mean SFR estimates for the host galaxy population.

Figure~\ref{fig:rate_sfr} shows clear evidence for a component of
SNe~Ia beyond that dependent on the integrated stellar mass. The
number of SNe~Ia expected in each SFR bin from the stellar mass is
smaller than actually observed; the majority of SNe~Ia in star-forming
galaxies appear to arise from more recently formed stars than from old
stars. Furthermore, the fraction of SNe~Ia from the ``A'' component
appears fairly constant with the SFR of the galaxy, with a range of
14\% to 21\%. We also test whether the dependence on the mean SFR of a
galaxy is linear by fitting the slope of the line, $n_{\mathrm{SFR}}$.
The best fit is $n_{\mathrm{SFR}}=0.84\pm0.06$ with a $\chi^2=1.31$
(the slope is $n_{\mathrm{SFR}}=0.81\pm0.08$ for the spectroscopically
confirmed SNe~Ia).  For comparison, a line with a slope of unity is
also shown ($\chi^2=2.34$). We estimate $B=4.1\pm0.7\times
10^{-4}$\,\Bunits\ using this approach and enforcing a linear
relationship.

\subsection{Bivariate fits}
\label{sec:bivariate-fits}

\begin{figure}
  \plotone{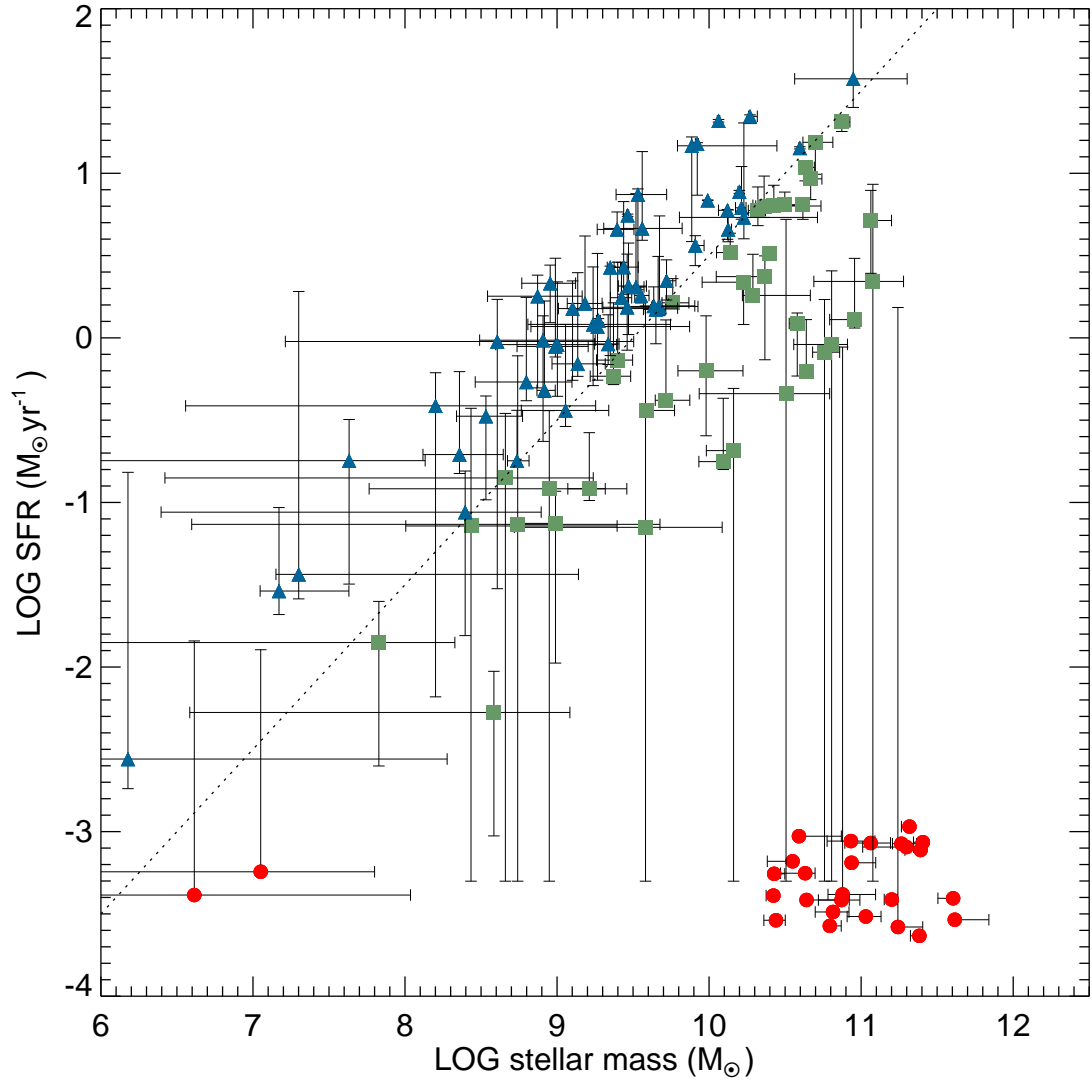}
\caption{
  The distribution of the SN~Ia host galaxies in the SFR-mass plane.
  Each galaxy is coded according to its assigned type. Passive
  galaxies are shown as circles (red), normal star-forming galaxies as
  squares (green), and vigorous star-formers as triangles (blue). The
  black diagonal dotted line shows the division in specific
  star-formation rate used to sub-divide those hosts that are
  star-forming. The passive galaxies (which have a zero SFR in our
  models) are assigned a random SFR centered on 0.005\mperyr\ for
  illustration purposes.
\label{fig:mass_sfr}
}
\end{figure}

\begin{figure}
\plotone{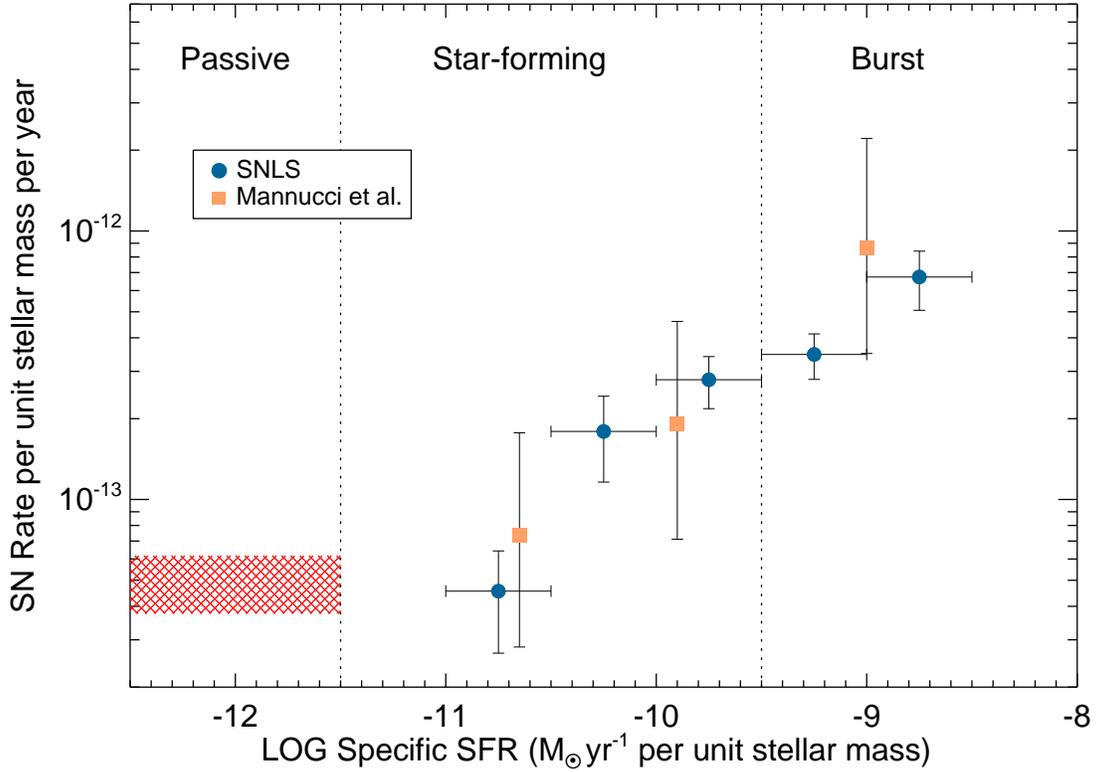}
\caption{
  The number of SNe~Ia per unit stellar mass as a function of the
  star-formation rate (SFR) per unit stellar mass of the host galaxy.
  Blue points represent SNLS data-points in star-forming galaxies. The
  red hashed area shows the number per unit stellar mass as measured
  in the SNLS passive galaxies (assigned zero SFR in our models).
  Shown for comparison is the evolution in SN~Ia rate to later-type
  galaxies observed locally by \citet{2005A&A...433..807M}, normalized
  to the SNLS rate in passive galaxies. The horizontal positioning of
  the Mannucci et al. data points are subject to a uncertainty when
  converting their galaxy types into specific SFRs. The vertical
  dotted lines show the division we use to classify the host galaxies
  into different types.
\label{fig:rate_ssfr}
}
\end{figure}

\begin{figure}
\plotone{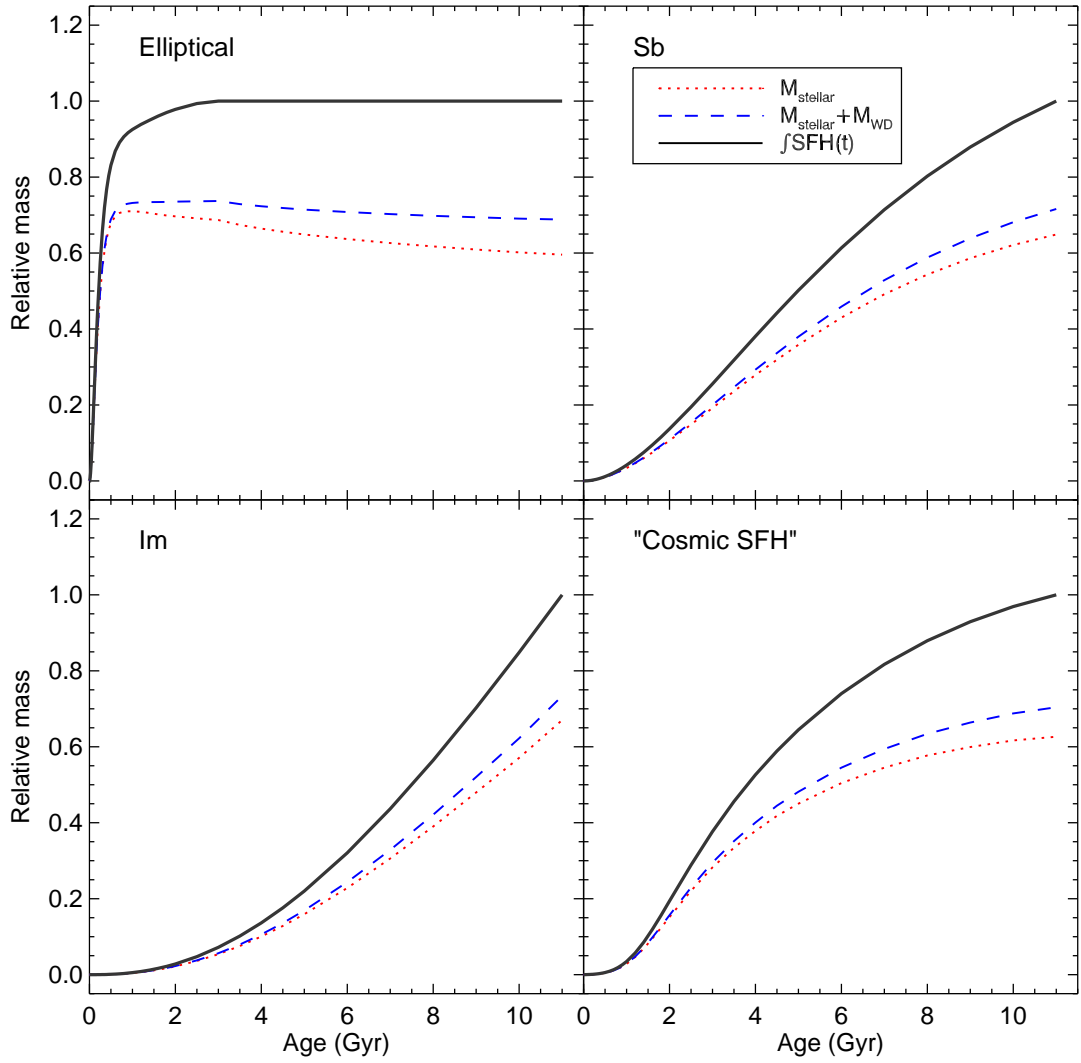}
\caption{
  The mass evolution, as calculated by P\'EGASE.2, for 3 different
  galaxy scenarios plus the cosmic SFH of \citet{2006astro.ph..1463H}. The
  different lines show the total mass in stars excluding compact
  objects (dotted line), the total mass in stars plus the mass in
  white dwarfs (dashed line), and the mass as calculated by simply
  integrating the SFH of that scenario (solid line).
\label{fig:mass-definitions}
}
\end{figure}

\begin{figure}
\plotone{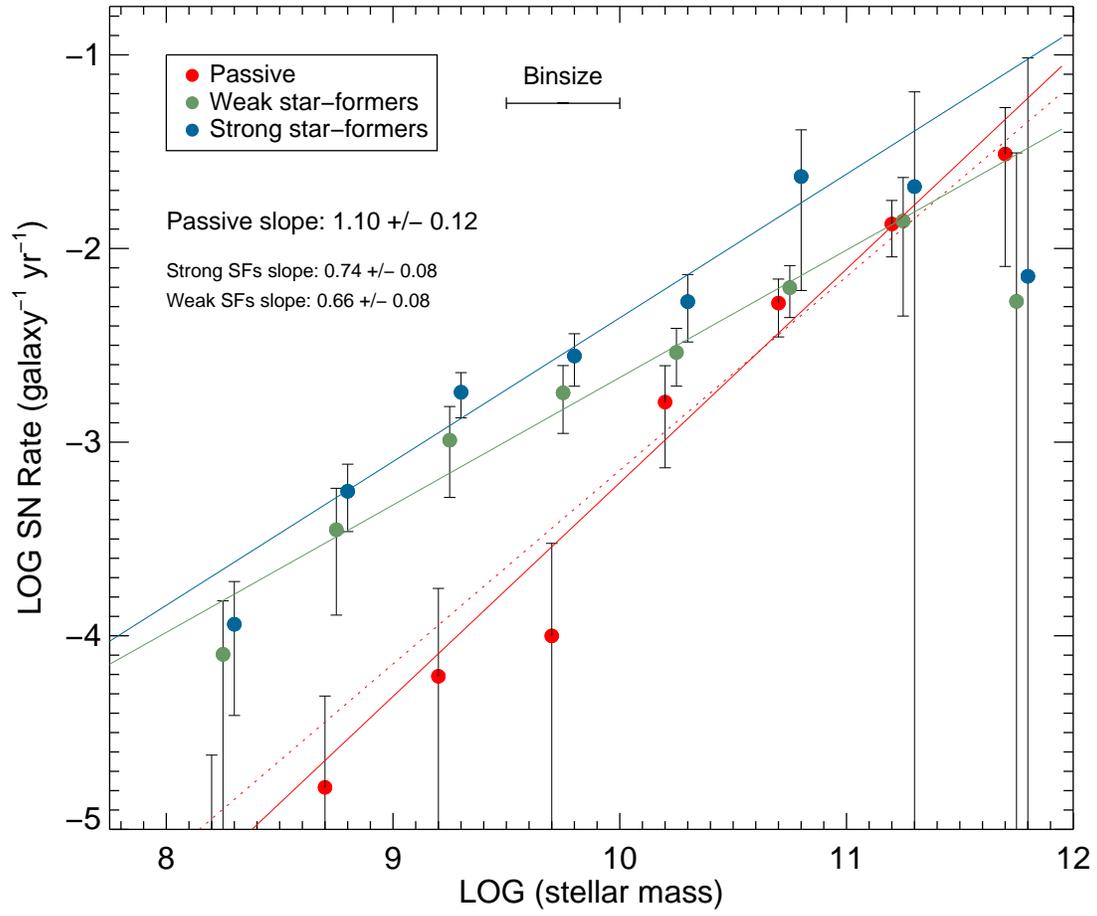}
\caption{
  The number of SNe~Ia per host galaxy as a function of host galaxy
  stellar mass. The three colors denote SNe~Ia in the three different
  types of host galaxy as partitioned by their specific star-formation
  rate. The best-fitting lines and slopes to each distribution are
  shown. For the passive hosts, a line of slope unity is also shown
  (dotted line).
\label{fig:rate_mass}
}
\end{figure}

\begin{figure}
\plotone{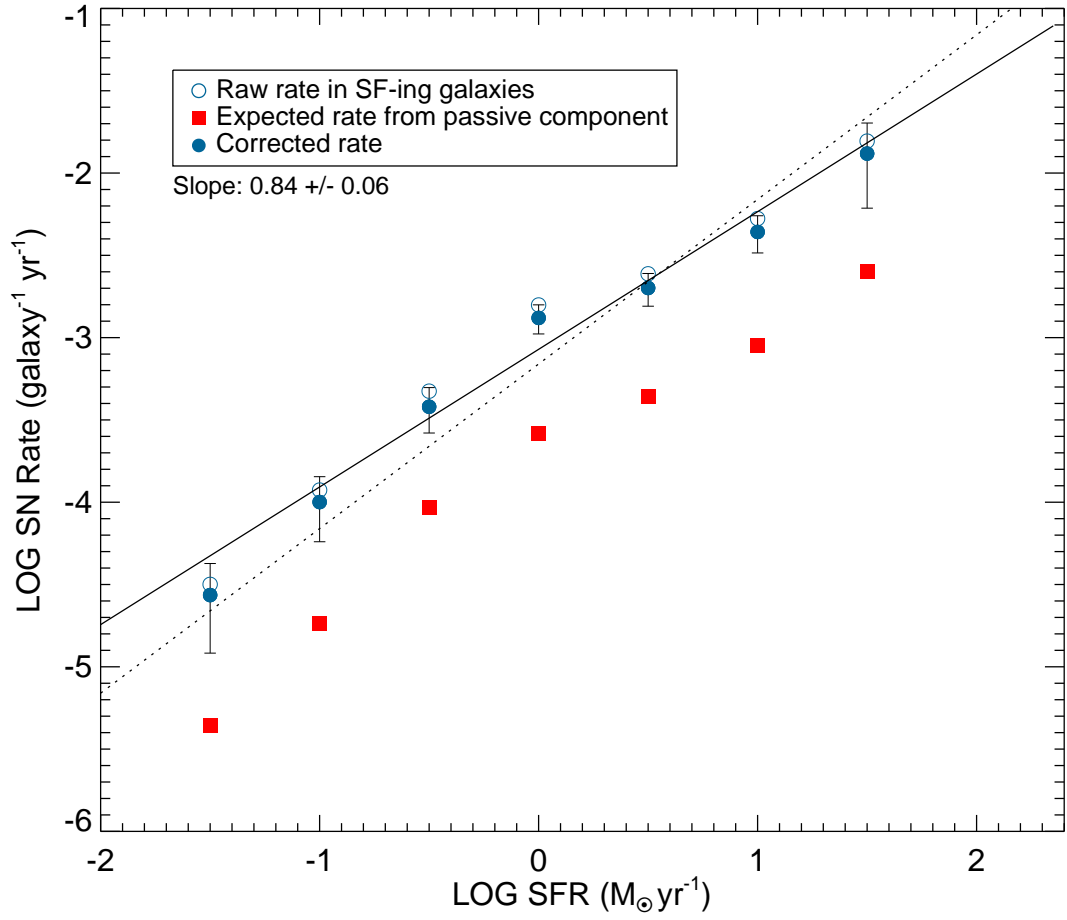}
\caption{
  The number of SNe~Ia per host galaxy as a function of host galaxy
  mean star-formation rate measured over a 0.5Gyr period (see text for
  details).  The open blue circles show the raw rate in star-forming
  galaxies, with the red filled squares showing the expected rate
  derived from the total stellar mass of field galaxies in each bin.
  The filled blue circles shows the number per galaxy after the
  component from the stellar mass is removed. The solid line shows the
  best-fit; the dotted line has a slope of unity.
\label{fig:rate_sfr}
}
\end{figure}

Though the last two sections provide a useful visualization of the
SN~Ia rate as a function of galaxy stellar mass and mean SFR, a more
sophisticated bivariate technique which fits $A$ and $B$
simultaneously across all galaxy types will result in more accurate
fit values, as the stellar masses and SFRs are partially correlated
(Figure~\ref{fig:mass_sfr}). We assume the relationship is a linear
function of only $A$ and $B$ (i.e. $n_{\mathrm{mass}}$ and
$n_{\mathrm{SFR}}$ are fixed), and perform a generalized linear
least-squares fit in the galaxy mass/SFR plane. This effectively fits
the two components of the SN~Ia rate simultaneously to the data. We
perform this fit by converting Figure~\ref{fig:mass_sfr}, the
distribution of SN~Ia host galaxies in the mass/SFR plane, into a
probability of a SN~Ia explosion in a given galaxy as a function of
mass and SFR. The probability is calculated by binning the host
distribution by galaxy mass and galaxy SFR, and dividing the number of
SN hosts in each bin by the (incompleteness corrected) number of field
galaxies similarly binned.  This results in a two dimensional mass/SFR
probability space which gives the likelihood of a SN~Ia explosion in a
galaxy as a function of both the galaxy stellar mass and mean SFR.

We fit two forms of equation~(\ref{eq:3}) to this probability
distribution: equation~(\ref{eq:3}) with $A$ and $B$ both free, and
the same function but with $B=0$ i.e. assuming the SN~Ia rate depends
only on host galaxy stellar mass (clearly the third alternative,
fixing $A=0$, makes no sense as plenty of SNe~Ia explode in
non-star-forming galaxies).  The addition of the $B$ term as a free
parameter ($B\ne0$) reduces the $\chi^2$ of the fit from 67 to 37 (60
degrees of freedom).  Performing an F-test, we find that the null
hypothesis that the extra $B$ term is \textit{not} needed is rejected
at $>$99.99\% probability.  The fit results when fitting for both $A$
and $B$ with $n_{\mathrm{mass}}=n_{\mathrm{SFR}}=1$ are
$A=5.3\pm1.1\times10^{-14}$ $(H_0/70)^2$\,\Aunits\ and
$B=3.9\pm0.7\times 10^{-4}$ $(H_0/70)^2$\,\Bunits\ (these errors are
statistical only). Using the formal covariance matrix of the fit, we
calculate a correlation coefficient between the input masses and SFRs
of -0.22, indicating only a small correlation.

One important feature of Figure~\ref{fig:rate_sfr} is the slight
non-linear relationship between \snrIa\ and SFR. Using the bivariate
fit method, we can also derive best-fit values for $n_{\mathrm{mass}}$
and $n_{\mathrm{SFR}}$ by examining the $\chi^2$ variation of the
best-fit as $n_{\mathrm{mass}}$ and $n_{\mathrm{SFR}}$ are varied in
the fitting function. We find minimum $\chi^2$ at
$n_{\mathrm{mass}}=1.00^{+0.11}_{-0.10}$
$n_{\mathrm{SFR}}=0.98^{+0.12}_{-0.11}$.  These are consistent though
more reliable than the estimates of $\S$~\ref{sec:sn-rate-mass} and
$\S$~\ref{sec:sn-rate-sfr}, and are completely consistent with
$n_{\mathrm{mass}}=n_{\mathrm{SFR}}=1$.

Clearly, in an analysis such as these systematic errors are likely to
be at least as large as the statistical ones quoted above. For
example, using different IMFs (e.g. those of
\citealt{1955ApJ...121..161S} or \citealt{1992A&A...265..499R}) can
vary the fit values of $A$ and $B$ by around 10-20\% of their value
compared to the IMF of \citet{2001MNRAS.322..231K}; using different
spectral libraries is likely to produce changes of a similar
magnitude. Furthermore, although dust is included in the P\'EGASE.2
models, no dust extinction correction is made to individual galaxies,
and systematic errors could clearly therefore be present in the SFR
estimates, especially when considering the existing evidence for
SFR-dependent extinction corrections
\citep{2001AJ....122..288H,2001ApJ...558...72S}.  Hence, the values of
$A$ and $B$ may change as more multi-wavelength data becomes available
for the host galaxies and better constrains the extinction properties
of the galaxies.

\subsection{Comparison to other results}
\label{sec:comp-other-results}

The fits of $\S$~\ref{sec:bivariate-fits} show that the SN~Ia rate in
galaxies has a linear dependence on both stellar mass and younger
stellar populations, which we parameterize via the galaxy SFR. We also
derived parameters $A$ and $B$ which allow us to relate the mass and
SFR of a galaxy to the probability of it hosting a SN~Ia. We can
compare our estimates of $A$ and $B$ with other similar numbers
published in the literature. 

\citet{2005A&A...433..807M} calculated the low-redshift SN~Ia rate per
unit mass as a function of host galaxy morphological type from the SN
sample of \citet{1999A&A...351..459C}.  In low-redshift E/S0 galaxies,
they find a rate of 3.83$^{+1.4}_{-1.2}\times10^{-14}$ \Aunits\ (in
our cosmology), which agrees with our determination
($A=5.3\pm1.1\times 10^{-14}$) within the error-bars.  Given the
different techniques for determining galaxy types (SED fitting versus
morphological typing), and the different search and efficiency
calculations, this is an encouraging level of agreement. Our
determination of $B$ ($3.9\pm0.7\times 10^{-4}$ \Bunits) is however
discrepant with values of SB05, who derive two different values,
$10^{+6}_{-5}\times 10^{-4}$ and $23^{+10}_{-10}\times 10^{-4}$ in our
units.  If we take the SNLS $z=0.47$ SN~Ia rate of
0.42$\times 10^{-4}$\,h$_{70}^3$\,SNe\,yr$^{-1}$\,Mpc$^{-3}$, and assume that all of
this rate is generated from the prompt component, with a
star-formation density at $z=0.47$ of 0.043 \mperyr\,Mpc$^{-3}$
\citep{2006astro.ph..1463H}, this implies an upper limit for $B$ of
$B\lesssim1.0\times 10^{-3}$ \Bunits, at the lower end of the SB05
assumed values.

We can also estimate the predicted rate of SNe~Ia in clusters from
ellipticals and compare to the observed cluster rate as a further
consistency check. As \textit{instantaneous} star-formation in massive
clusters can be suppressed relative to field galaxies
\citep{2001ApJ...549..820C}, cluster SNe~Ia are likely to be dominated
by events from the old component.  \citet*{2002MNRAS.332...37G}
measure a cluster SN~Ia rate of 0.392h$^2_{70}$\,SNu\footnote{1
  ``SNu'' is one SN per century per 10$^{10}$ stellar $B$-band
  luminosities}.  Assuming a typical elliptical mass-to-light ratio in
$B$-band of 5-10, this converts to a SN~Ia rate in clusters of
3.9-7.8\,\Aunits, in excellent agreement with our $A$ parameter
determination.

We also check that the $A$ and $B$ values derived here are consistent
with measured volumetric SN~Ia rates. We use the cosmic SFH of
\citet{2006astro.ph..1463H} which gives the SFR of the universe as a
function of redshift. Integrating this SFR gives the mass as a
function of redshift, which in conjunction with our $A$ and $B$ values
will predict the SN~Ia rate. Care must be taken that a similar
definition of mass is used here as is used when we derive $A$ i.e. the
mass of stars that have died must be subtracted from the integral (see
Figure~\ref{fig:mass-definitions} for the difference this correction
makes in the mass evolution of the cosmic SFH). To do this, we use the
SFH of \citet{2006astro.ph..1463H} as an input to P\'EGASE.2 to
calculate the total mass in stars as a function of redshift. We then
use the $A$ and $B$ values from above to calculate the volumetric
SN~Ia rate (Figure~\ref{fig:rate_z}). The relative contributions of
the two components evolves strongly with time. The young component
provides $\sim$10-20\% of all SNe~Ia at $z=0$, rising to $\sim85$\% at
$z=2$; the exact ratios depend strongly on the assumed cosmic SFH.
Qualitatively similar trends are also predicted by
\citet{2005astro.ph.10315M}, based on completely different analysis
and fitting techniques. This first fraction is quite low; we note that
assuming that the likelihood of the old component decreases with age
rather than remains constant as in eqn.~(\ref{eq:3}) would result in a
larger fraction of SNe being generated from the prompt component at
$z=0$, perhaps illustrating a limitation of the ``A+B'' model.

The agreement with the volumetric rates is remarkably good, given the
SFH used in the calculation \citep{2006astro.ph..1463H} is completely
independent of the derivation of $A$ and $B$. All of the published
rates are statistically consistent at the 2-$\sigma$ level with the
simple ``prompt+old'' model parameters derived here. The most deviant
point is the $\simeq1$ rate from \citet{2004ApJ...613..189D}, but even
this value differs from the model at only 1.8\,$\sigma$. The $A$ and
$B$ model used here predicts a shallow redshift evolution of the SN~Ia
rate, one that does not evolve as fast as the cosmic SFH.  Of course,
the distribution of the two types will vary considerably with local
galaxy density. In low-mass star-forming dwarf field galaxies, for
example, the prompt component will be the source for essentially all
SNe~Ia.  Yet, a substantial fraction of SNe~Ia will always occur in
E/S0 galaxies (and perhaps the bulges of spiral galaxies) and in
clusters due the large amount of old stellar mass locked up in these
systems.

\section{Supernova properties as a function of stellar populations}
\label{sec:supern-prop-as}

The results of $\S$~\ref{sec:supernova-rate} show that the rate of
SNe~Ia in a given galaxy depends on both the evolved (old) stellar
mass of that system, and on the young stellar content, parameterized
via the SFR. Indeed, a simple mass-scaling -- assuming that the SN~Ia
rate is a simple function of galaxy stellar mass -- is unable to
reproduce the trends that we see. These results could have important
implications for the use of SNe~Ia as precise cosmological probes.  At
the very least, as discussed by SB05 and \citet{2005astro.ph.10315M},
there must be a broad range of delay times between a binary system
formation and SN explosion.  Whether this is interpreted as being due
to a single channel to a SN~Ia (i.e.  perhaps single degenerate
\textit{or} double degenerate), or via a two channel path to a SN~Ia
(i.e. perhaps single degenerate \textit{and} double degenerate), a
distinct possibility is that the average properties of these
``prompt'' and ``old'' SN~Ia populations could differ. In this
section, we search for SN~Ia properties that correlate with
environment, or more particularly star-formation activity (or the mean
age of the stellar population) in their host galaxies.

One key observable affecting the utility of SNe~Ia as cosmological
probes is the light-curve shape/luminosity relationship
\citep{1993ApJ...413L.105P}, which corrects the luminosity of SNe~Ia
according to the width of their light-curves -- the so called
``brighter--slower'' relationship. In this paper, we parameterize the
SN~Ia light-curves using the stretch parameter $s$
\citep[e.g.][]{1997ApJ...483..565P}, which linearly stretches or
contracts the time axis of a template SN~Ia light curve around the
time of maximum light to best-fit the observed light curve of the SN
being fit. At low redshift, a trend of SN~Ia absolute luminosity -- or
equivalently, light-curve width cast in terms of \dmB, the decline in
$B$-band magnitudes 15 days after maximum light -- versus galaxy
morphological type has been observed \citep[][see
\citealt{2005ApJ...634..210G} for a recent
compilation]{1995AJ....109....1H,1996AJ....112.2398H,1999AJ....117..707R,2000AJ....120.1479H}.
These authors show that brighter SNe~Ia (or more precisely
high-stretch/low-\dmB) are preferentially located in late-type
(younger) galaxies.

Here, we compare the low-redshift distribution to that seen at high
redshift in the current SNLS sample (Figure~\ref{fig:stretch}).  We
use the specific SFR classification of the SNLS host galaxies at high
redshift, and take classifications for the low-redshift SNe hosts from
either the literature
\citep{1999AJ....117..707R,2000AJ....120.1479H,2002PASP..114..820V} or
from the NASA/IPAC Extragalactic Database (NED), simplifying onto a
elliptical (E/S0), early-type (Sa to Sbc) and late-type (Sc and later)
classification scheme. We refit the published light-curves for the
low-redshift SNe~Ia with the same method as the high-redshift SNe~Ia
to ensure a consistent definition of ``stretch'' between low and high
redshift.

The trend in Figure~\ref{fig:stretch} across galaxy type is clear --
passive galaxies with a small (or zero) specific SFR tend to host
lower stretch SNe. Though the low-$z$ trend was already well-known,
this is the first time the effect has been seen at high-$z$
\citep[e.g.][]{2003MNRAS.340.1057S}, presumably because the SNLS
probes a wider range in stretch than previous high-$z$ surveys,
particular at lower stretches (fainter SNe). Kolmogorov-Smirnov (K-S)
tests can be used to give some statistical guide as to whether these
various stretch distributions are similar; these are summarized in
Table~\ref{tab:kstests}. We show that at high redshift, the stretch
distributions between SNe in passive galaxies versus those in both
types of star-forming galaxies are different with $>$98\% confidence.
We also show that the stretch distributions for spirals at low-$z$ and
star-forming galaxies at high-$z$ are essentially identical. 

The distributions in ellipticals at low-$z$ versus passive galaxies at
high-$z$ are formally different at the 90\% level, though this is
caused entirely by the lack of very low stretch high-$z$ SNe
($s\sim0.7$). One possibility is that the delay times for the lowest
stretch SNe~Ia are long enough to preclude their existence at
high redshift \citep{2001ApJ...554L.193H}. Another is that as these
SNe are the fainter subsample, this absence could be attributed to
selection effects. 

In general, selection effects are unlikely to be responsible for the
other trends that we observe. The main trend concerns an absence of
high-stretch (i.e. brighter) SNe~Ia in passive systems at high
redshift; a selection or Malmquist bias in passive systems would lead
to the opposite trend, i.e. a decrease in the lower-stretch (fainter)
SNe~Ia in these systems. Furthermore, the trends remain unchanged if the
sample is culled at $z=0.6$ (where Malmquist effects will be smaller)
rather than $z=0.75$.

These trends seem highly suggestive of differing properties between
prompt and older SNe~Ia. The passive galaxies, with a SFR=0 and an
average older stellar population, host events with average stretches
differing from those prompt SNe~Ia found in galaxies comprised of
younger stellar populations. Of course, galaxies with an average young
stellar population are still likely to host low-stretch SNe~Ia due to
the older stellar populations which must be present in all but the
youngest starburst galaxies, though the reverse is less likely to be
true. We can attempt to identify this low stretch population in
Figure~\ref{fig:stretch} using a simple mass-scaling technique in the
context of the single degenerate progenitor system scenario.

We sum the total stellar mass in all the field passive galaxies and in
all the star-forming galaxies. We then scale the stretch distribution
in the old galaxies by the ratio of these two total masses, and
overplot on the stretch distributions of SNe~Ia in the younger hosts.
Subtracting this scaled population leaves just the residual stretch
distribution remaining (Figure~\ref{fig:stretch_residual}). Clearly,
other ratios could be used to scale the passive distribution prior to
subtraction.  We found that the residual stretch distribution was
similar when using various different mass ranges (e.g., 0.8\,\msun\ to
3.0\,\msun) to calculate the scaling ratios.

The resulting subtraction is remarkably clean given the simple
assumption that was made when scaling the low-stretch SN distribution.
The low-stretch distribution in passive galaxies seems able to
reproduce the distribution of low-stretch SNe~Ia in star-forming
galaxies, within the statistical errors. This seems to provide
supporting evidence that not only is age a key parameter driving the
SN~Ia rate (see $\S$~\ref{sec:supernova-rate}), but that it also
provides a physical parameter partially controlling the observed SN to
SN stretch variation. 

\section{Conclusions}
\label{sec:conclusions}

In this paper, we have examined the rates and properties of
high-redshift Type Ia supernovae (SNe~Ia) discovered via the Supernova
Legacy Survey (SNLS) as a function of the stellar mass and
star-formation in their host galaxies. Our principal findings are:

\begin{enumerate}
\item The SN~Ia rate per unit stellar mass is a strong function of
  host galaxy specific star-formation rate. More strongly star-forming
  (later-type) galaxies host around 10 times as many SNe~Ia per unit
  mass than do passive galaxies with a zero star-formation rate,
  similar to trends observed in the local Universe by
  \citet{2005A&A...433..807M}.
\item The number of SNe~Ia per galaxy in passive galaxies closely
  tracks the stellar mass of the host system with a linear
  relationship.  Though a relationship between the number of SNe per
  galaxy and galaxy mass is also seen in star-forming galaxies, it is
  not a simple linear dependence ($n=\simeq0.7\pm0.08$), with an
  excess of SNe~Ia in low-mass star-forming galaxies when compared to
  that in non-star-forming galaxies.
\item We find a clear relationship between the number of SNe~Ia per
  galaxy and the galaxy mean star-formation rate averaged over the
  last 0.5\,Gyr, even after removing any signal derived from the
  SN/host mass relationship found above.  More strongly star-forming
  systems host more SNe~Ia per galaxy than lower star-formation rate
  systems, with a best-fitting slope of
  $n_{\mathrm{SFR}}=0.98^{+0.12}_{-0.11}$.
\item By approximating the SN~Ia rate as a bivariate linear function
  of host galaxy stellar mass ($M$) and host galaxy mean SFR averaged
  over the last 0.5\,Gyr (${\dot M_{new}}$), we find that the SN~Ia
  rate in a galaxy is well represented by $\snrIa(t)=$ $AM(t)+B{\dot
    M_{new}}(t)$, with $A=5.3\pm1.1\times10^{-14}$
  $(H_0/70)^2$\,\Aunits, and $B=3.9\pm0.7\times 10^{-4}$
  $(H_0/70)^2$\,\Bunits.
\item We demonstrate, for the first time at high redshift, a
  relationship between star-formation in a host galaxy and the SN
  light-curve width (stretch). We find that high-stretch (brighter)
  SNe~Ia are exclusively hosted by star-forming galaxies, while
  non-star-forming galaxies only host low-stretch SNe~Ia. We show that
  the SN stretch distributions in low redshift and high redshift
  spirals are statistically identical at 95\% confidence, with the
  distribution in ellipticals identical at low and high redshift
  except for the lowest-stretch SNe~Ia which are not present at
  high redshift.
\item We show that the total SN~Ia stretch distribution in
  high-redshift star-forming galaxies can be well represented by a
  combination of a high-stretch component and a low-stretch component
  equivalent to the low-stretch distribution in passive galaxies,
  scaled by the ratio of the total mass in passive galaxies to that in
  star-forming galaxies.  The indication is that not only can SNe~Ia
  be generated from both old and young progenitor systems, but there
  is a systematic difference in the mean light-curve properties of the
  two components.
\end{enumerate}

\begin{figure}
\plotone{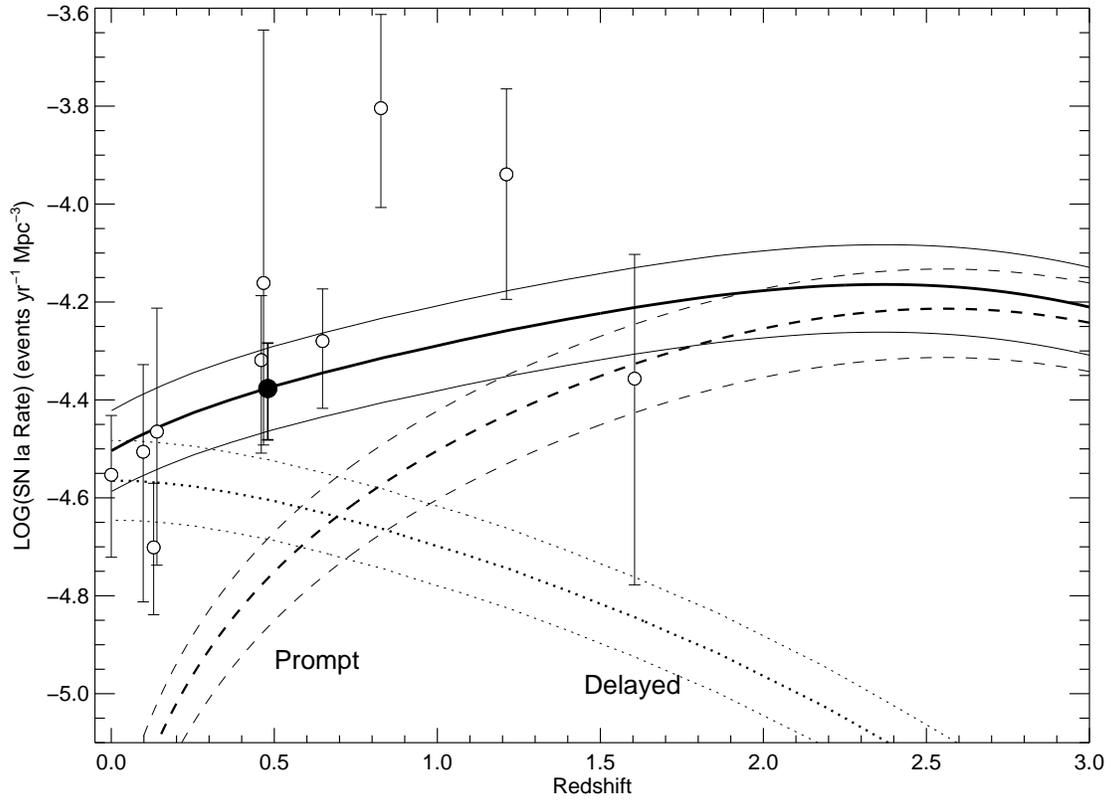}
\caption{
  The predicted volumetric SN~Ia rate as a function of redshift based
  on the $A$ and $B$ values from the bivariate fit of
  $\S$~\ref{sec:bivariate-fits}. The dotted lines denotes the $A$
  ``old'' component (plus limits from the statistical errors), and the
  dashed line the $B$ ``prompt'' component. The solid line shows the
  sum of the two. The filled circle is the $z=0.47$ SNLS determination
  of the SN~Ia rate of \citet{2006neill}. Open circles represent other
  SN~Ia rate determinations from \citet{1999A&A...351..459C},
  \citet{2000A&A...362..419H}, \citet{2002ApJ...577..120P},
  \citet{2003ApJ...594....1T}, \citet{2004ApJ...613..189D} and
  \citet{2004A&A...423..881B}. We conservatively show the statistical
  and systematic error-bars added in quadrature where both are given
  in these papers. The star-formation history of
  \citet{2006astro.ph..1463H} is assumed.
\label{fig:rate_z}
}
\end{figure}

\begin{figure}
\plottwo{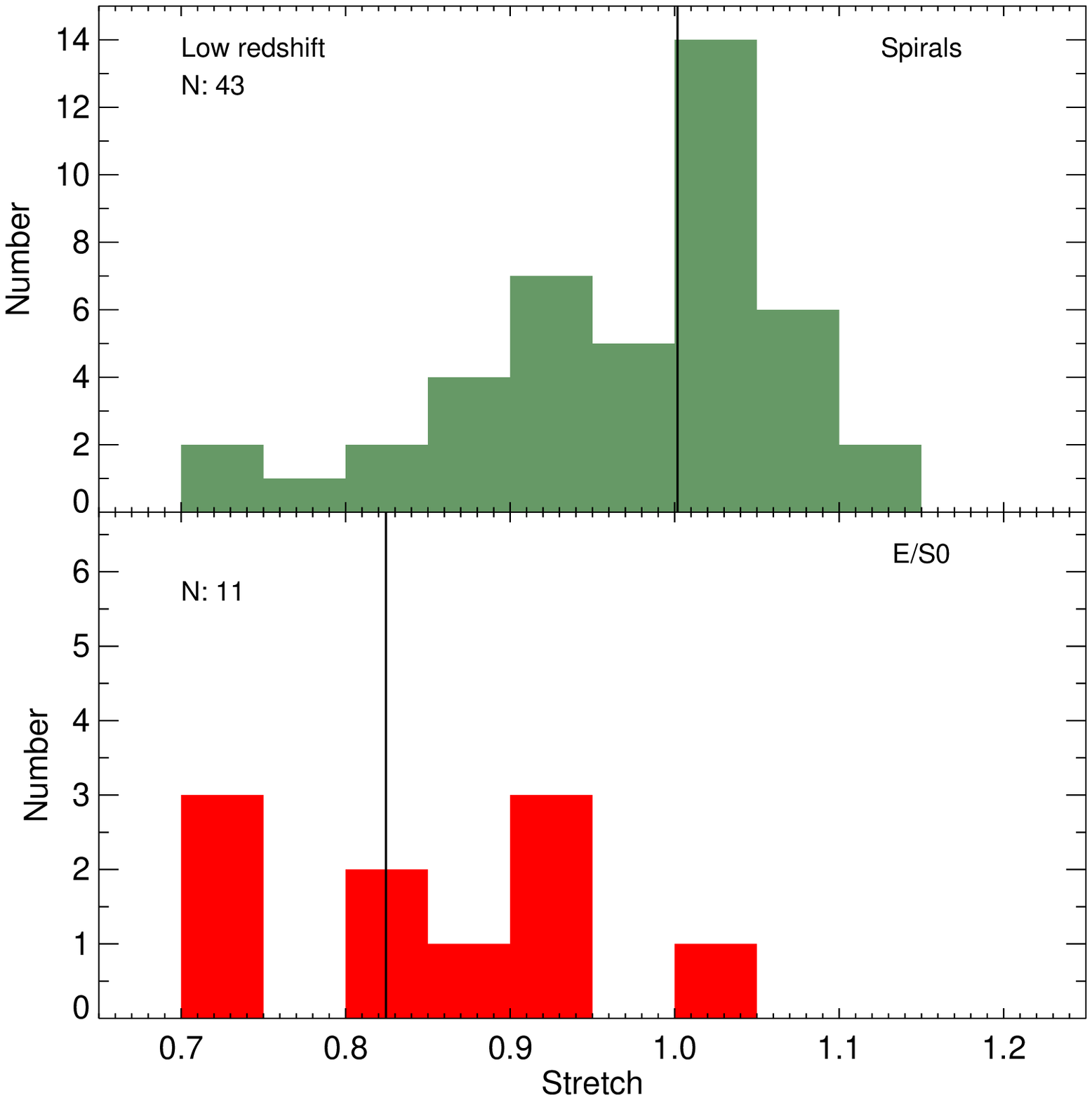}{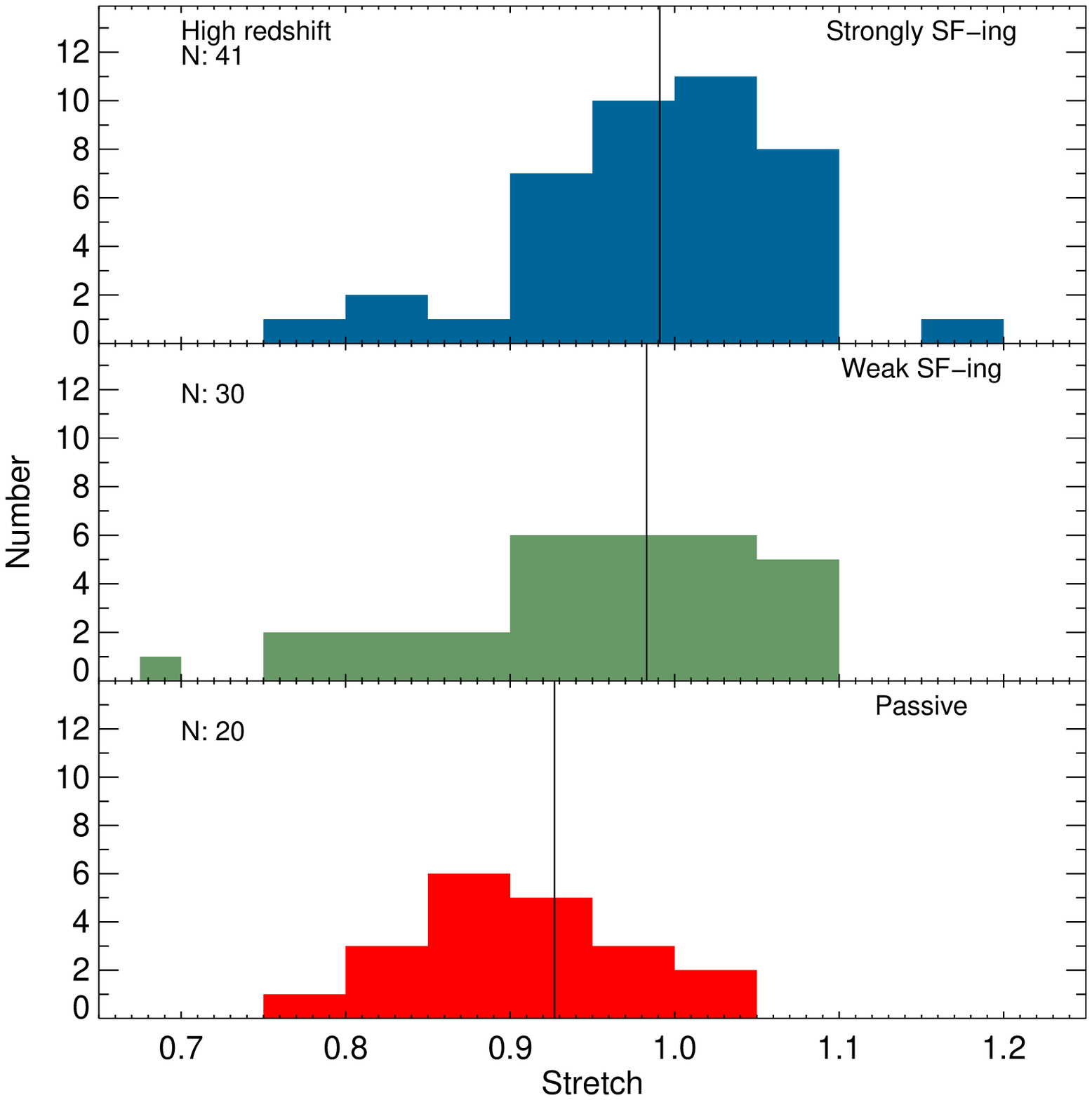}
\caption{
  The distribution of the SN~Ia light-curve shape parameter
  ``stretch'', separated according to the specific star-formation rate
  of the host galaxy. The left-hand plot shows low-redshift SNe~Ia,
  the right-hand plot shows high-redshift SNLS SNe~Ia. The typical
  precision on the stretch measure is $\pm 0.01-0.02$, i.e.  smaller
  than the bin width of the histograms. LEFT: The top panel shows
  galaxies morphologically classified as spirals, the lower panel
  shows those SNe in elliptical or S0 galaxies. RIGHT: The top panel
  shows galaxies with a specific SFR (sSFR) of $\log (\mathrm{sSFR})>
  -9.5$, the middle panel galaxies with $-12.0 \le \log
  (\mathrm{sSFR}) \le -9.5$, and the lower panel $\log
  (\mathrm{sSFR})< -12.0$. The vertical lines show the positions of
  the median stretch in each histogram.
\label{fig:stretch}
}
\end{figure}

\begin{figure}
\plotone{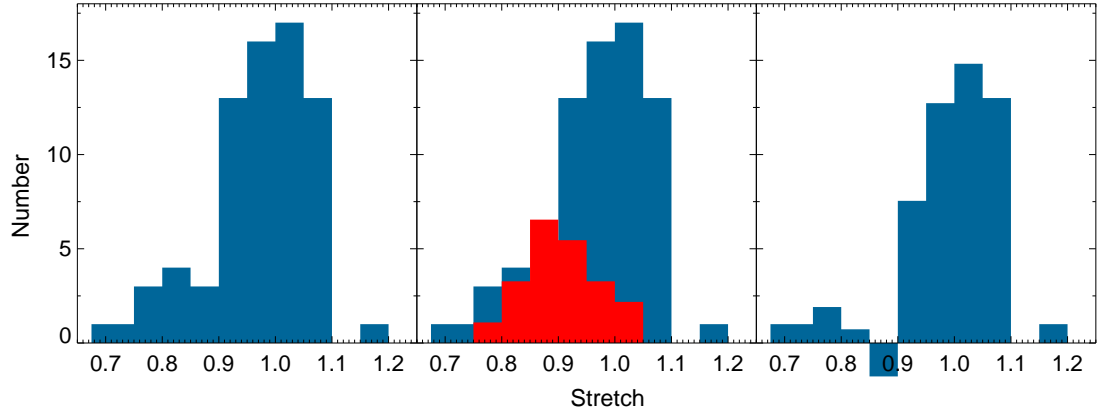}
\caption{
  The effect of subtracting the SN~Ia stretch distribution in passive
  galaxies from the distribution in star-forming galaxies.  The
  left-hand panel shows the distribution in all star-forming galaxies,
  the middle panel shows the passive galaxy distribution over-plotted
  scaled by the ratio of the total mass in passive galaxies to the
  total mass in star-formers. The right-hand panel shows the remaining
  distribution after subtracting the scaled passive galaxy
  distribution.
\label{fig:stretch_residual}
}
\end{figure}

These conclusions could have implications for the use of SNe~Ia to
determine cosmological parameters. As the cosmic star-formation rate
density shows sharp evolution as a function of redshift, the relative
mix of the two SN~Ia components will change correspondingly (see
Figure~\ref{fig:rate_z}), \textit{assuming the efficiency of
  generating a SN~Ia from a given progenitor scenario is invariant
  with redshift}. The prompt component should supply the dominant
fraction of observed SNe~Ia at high redshift, with the old component
producing a larger fraction at low redshift. The cross-over point,
where the contributions from the two components is the same, is around
$z\sim0.5-0.9$ assuming the \citet{2006astro.ph..1463H} star-formation
history.

There are at least two direct implications for the use of SNe~Ia as
cosmological probes. The first is that SNe~Ia should be found even at
very high redshift ($z\sim3-4$ and above). While SN~Ia models with
significant delay-times would preclude the existence of SN~Ia when the
Universe was so young ($t~\simeq1-2$\,Gyr), the delay-time for the
prompt component of SNe~Ia is very short. The implication is that
SNe~Ia could therefore be used as cosmological probes up until the
highest redshifts at which stars are being formed. SNe~Ia at $z=4-5$
would be within easy range of future facilities such as the James Webb
Space Telescope (\textit{JWST}) and the various proposed 30-100m class
ground-based telescopes.

The second implication is that an excellent understanding of the
light-curve shape/luminosity correction in different environments will
be essential to fully exploit SNe~Ia in measurements of $<w>$, and, in
particular, for the more sensitive task of measuring any variation of
$w$ with redshift. The study and classification of SN~Ia environment,
using, for example, similar techniques to those presented here, may
become as important a part of determining future cosmological
constraints as measuring the light-curves of the SNe~Ia themselves.
Surveys which routinely obtain detailed information on the environment
of all confirmed SNe detected will be ideally placed to perform
studies of this nature.

\acknowledgments

The SNLS collaboration gratefully acknowledges the assistance of
Pierre Martin and the CFHT Queued Service Observations team.
Jean-Charles Cuillandre and Kanoa Withington were also indispensable
in making possible real-time data reduction at CFHT. We thank Lars
Bildsten and Evan Scannapieco for useful discussions, and Andrew
Hopkins for providing a copy of Hopkins \& Beacom (2006) prior to
submission. Based on observations obtained with MegaPrime/MegaCam, a
joint project of CFHT and CEA/DAPNIA, at the Canada-France-Hawaii
Telescope (CFHT) which is operated by the National Research Council
(NRC) of Canada, the Institut National des Sciences de l'Univers of
the Centre National de la Recherche Scientifique (CNRS) of France, and
the University of Hawaii. This work is based in part on data products
produced at the Canadian Astronomy Data Centre as part of the CFHT
Legacy Survey, a collaborative project of NRC and CNRS.  Canadian
collaboration members acknowledge support from NSERC and CIAR; French
collaboration members from CNRS/IN2P3, CNRS/INSU and CEA. This
research has made use of the NASA/IPAC Extragalactic Database (NED)
which is operated by the Jet Propulsion Laboratory, California
Institute of Technology, under contract with the National Aeronautics
and Space Administration. The views expressed in this article are
those of the authors and do not reflect the official policy or
position of the United States Air Force, Department of Defense, or the
U.S. Government.

%%{\it Facilities:} \facility{CFHT}, \facility{Gemini}, \facility{VLT}, \facility{Keck}.

\bibliographystyle{apj}

\begin{deluxetable}{lllllll}
\tablecaption{Input and recovered masses and SFRs for our simulated galaxies}
\tablehead{\colhead{} & \multicolumn{3}{c}{Redshift known} & \multicolumn{3}{c}{Redshift free}\\ \colhead{Quantity} & \colhead{Offset\tablenotemark{a}} & \colhead{St. dev.} & \colhead{Fraction of outliers} & \colhead{Offset} & \colhead{St. dev.} & \colhead{Fraction of outliers}}
\startdata
Mass                               & -0.11  & 0.31 & 0.042 & -0.02 & 0.29 & 0.101 \\
$\overline{\mathrm{SFR}}$(0.05Gyr) &  0.11  & 0.51 & 0.003 & 0.39 & 0.51  & 0.032 \\
$\overline{\mathrm{SFR}}$(0.10Gyr) &  0.04  & 0.47 & 0.003 & 0.32 & 0.55  & 0.026 \\
$\overline{\mathrm{SFR}}$(0.20Gyr) & -0.04  & 0.43 & 0.006 & 0.23 & 0.55  & 0.023 \\
$\overline{\mathrm{SFR}}$(0.50Gyr) & -0.10  & 0.46 & 0.003 & 0.11 & 0.59  & 0.026 \\
$\overline{\mathrm{SFR}}$(1.00Gyr) & -0.14  & 0.52 & 0.031 & 0.02 & 0.58  & 0.076 \\
\enddata
\tablenotetext{a}{All numerical values are in dex units}
\label{tab:simulations}
\end{deluxetable}

\begin{deluxetable}{lcccccc}
\tablecaption{Breakdown of candidate numbers and efficiencies}
\tablehead{ \colhead{Field$^{\tablenotemark{a}}$} & \colhead{Area} & \colhead{N$_{\mathrm{spec}}$$^{\tablenotemark{b}}$} & \colhead{N$_{\mathrm{phot}}$$^{\tablenotemark{c}}$} & \multicolumn{3}{c}{Efficiency}\\ & \colhead{sq. deg.} & & & \colhead{0.2-0.4} &\colhead{ 0.4-0.6} & \colhead{0.6-0.75}}
\startdata
D1 & 1.024 & 25 & 5  & 0.314 & 0.297 & 0.241\\
D2 & 1.026 & 27 & 4  & 0.223 & 0.218 & 0.181\\
D3 & 1.029 & 28 & 11 & 0.331 & 0.310 & 0.248\\
D4 & 1.027 & 20 & 4  & 0.307 & 0.317 & 0.274\\
\enddata
\tablenotetext{a}{Each field comprises SNe from two years of observing}
\tablenotetext{b}{Spectroscopically confirmed SNe~Ia with a confidence index from 3 to 5 \citep{2005ApJ...634.1190H}}
\tablenotetext{c}{SNe~Ia with either a spectroscopic redshift but no definitive type, or solely a photometric redshift (see $\S$~\ref{sec:sn-incompl-corr})}
\label{tab:eff_info}
\end{deluxetable}

\begin{deluxetable}{lcccc}
\tablecaption{Breakdown of photometric typing culls}
\tablehead{ \colhead{Criterium} & \multicolumn{2}{c}{N$_{\mathrm{spec}}$} & \multicolumn{2}{c}{N$_{\mathrm{phot}}$} \\ \colhead{} &\colhead{} &\colhead{} &\colhead{} &\colhead{}}
\startdata
Number over 0.2$<$$z$$<$0.75            & 116 &    & 286 &    \\
Number rejected on light-curve coverage &     & 16 &     & 80 \\
Number rejected on \zspec\              &     & \nodata   &     & 2  \\
Number rejected on stretch              &     & \nodata   &     & 70 \\
Number rejected on $\chi^2$             &     & \nodata   &     & 75 \\
Number rejected on $g'$ dispersion      &     & \nodata   &     & 35 \\
Number remaining                        & 100 &    & 24  &    \\
\enddata
\label{tab:culls}
\end{deluxetable}

\begin{deluxetable}{lc}
\tablecaption{K-S tests on the stretch distributions at low and high redshift}
\tablehead{ \colhead{K-S test sample} & \colhead{Probability that distribution}\\ \colhead{} & \colhead{is the same}}
\startdata
Low-$z$  E/S0    $\rightarrow$ High-$z$ passive  & 10\% \\
Low-$z$  spirals $\rightarrow$ High-$z$ SFR$>$0  & 95\% \\
High-$z$ passive $\rightarrow$ High-$z$ weak$^{\tablenotemark{a}}$ SF-ing & 2\% \\
High-$z$ passive $\rightarrow$ High-$z$ strong SF-ing & 0.2\% \\
High-$z$ passive $\rightarrow$ High-$z$ all SF-ing & 0.9\% \\
High-$z$ weak SF-ing $\rightarrow$ High-$z$ strong SF-ing & 53\% \\
\enddata

\tablenotetext{a}{The weak and strong star-forming galaxies are
  divided based on their specific star-formation rate; see
  $\S$~\ref{sec:supernova-rate}}

\label{tab:kstests}
\end{deluxetable}

\clearpage

\clearpage

\clearpage

\clearpage

\clearpage

\clearpage

\clearpage

\clearpage

\clearpage

\clearpage

\clearpage

\clearpage

\end{document}